**Reformulating Scalar-Tensor Field Theories**

**as Scalar-Scalar Field Theories**

**Using Lorentzian Cofinsler Spaces**

**by**

**Gregory W. Horndeski**
**2814 Calle Dulcinea**
**Santa Fe, NM 87505-6425**
**email:**
**horndeskimath@gmail.com**

**September 5, 2020**

# ABSTRACT

In this paper I shall show how the notions of Finsler geometry can be used to construct a similar geometry using a scalar field, f, on the cotangent bundle of a differentiable manifold M. This will enable me to use the second vertical derivatives of f, along with the differential of a scalar field $\varphi$ on M, to construct a Lorentzian metric on M that depends upon $\varphi$. f will be chosen so that the resultant metric on M has the form of a FLRW metric, with the t equal constant slices being flat. When the Horndeski Lagrangians are evaluated for this choice of geometry the quartic and quintic Lagrangians are third order in $\varphi$, but reduce to non-degenerate second-order Lagrangians plus a divergence. Upon varying $\varphi$ in these "scalarized" Horndeski Lagrangians, equations will be obtained which admit self-inflating universe solutions, provided the coefficient functions in the Lagrangians are chosen suitably. This approach is also used to study solutions of the most general conformally invariant scalar-tensor field theory which is flat space compatible (i.e., such that the Lagrangians are well-defined when either the space is flat or the scalar field is constant). There too the coefficient functions can be chosen to give self-inflating universes. Arguments will be presented to show that it may be possible to construct model Universes that begin explosively, and then settle down to period of much quieter acceleration, which can be followed by a collapse to its original, pre-expansion, state.

**Table of Contents**

## Section 1: Introduction

I shall begin with a few remarks which are intended to help explain why I considered using ideas from Finsler geometry to study scalar-tensor field theories.

When a layman asks me what a scalar field is I usually answer with an example, by telling them that the temperature at each point in (say) this room, at each instant of time is a good example of a scalar field. However, that is not really true. After all temperature is a measure of energy, and such measurements are observer dependent–but I wish to spare them the confusing details. However, if I said the temperature at each point in the room at each instant of time measured by observers at rest with respect to me, then I would have a valid scalar field.

The important thing to be gleaned from the temperature example is, as is well-known, energy measurements are observer dependent. I shall now show how this can be used to construct an energy scalar field on the subbundle, TLUM, of time-like unit vectors on a Lorentzian spacetime $V_4 = (M,g)$, with signature $(-, +, +, +)$. This in turn will get us thinking about Finsler Spaces as suitable venues for studying gravitation.

Let P be a point of M, and let $v_P$ be the unit TL (:= timelike) tangent vector to the world line of an observer, $O_P$, at P. If $\mu_P \in T_P M$ is the energy-momentum vector of a particle passing through P, then the energy of that particle with respect to $O_P$ is $E(\mu_P ,v_P) := -g(\mu_P, v_P)$. We can now define the energy with respect to $O_P$ of all particles

passing through P by

$$E(v_P) := -\sum_{\substack{\text{all particles} \\ \text{passing through P}}} g(\mu_P, v_P) .$$  Eq.1.1

The function $E: TLUM \rightarrow \mathbb{R}$ is closely related to a scalar field $\mathscr{E}$ on TLM, the bundle of TL vectors on M, which is an open subbundle of TM. If $v \in TLM$ then

$$\mathscr{E}(v) := E((-g(v,v))^{-1/2} v) .$$  Eq.1.2

In general reltivity it is more convenient to deal with the energy density of matter fields, which is obtained from the energy-momentum tensor, T, of those fields.  T is an (0,2) tensor field, and gives rise to an energy density scalar function $\mathscr{E}_T$ on TLM, which is defined by

$$\mathscr{E}_T(v) := -T(v,v)/g(v,v) .$$  Eq.1.3

Since energy resides as a scalar field in TM, it seems reasonable to try to reformulate gravity as a field theory on TLM or TLUM.  This is where Lorentzian Finsler Geometry can be useful.  Such geometries are relatively new in comparison to Finsler Geometries which arose in the early 1900's , Finsler [1].  Excellent treatises on Finsler Geometry are provided by Rund [2], and Chern, *et al* [3]. I shall not bother giving the complete definition of Finsler Spaces, but a few cursory remarks are in order.  I shall be a bit more precise when we get to pseudo-Finsler spaces.

If we have a curve c = c(t), $t_i \le t \le t_f$, in a Riemannian Manifold $V_n = (M,g)$, we define the length of c by

$$L(c) := \int_{t_i}^{t_f} [g(c',c')]^{1/2} dt ,$$

where c' is the tangent vector to c, which can be viewed as a curve in TM. If $\mathbf{c}=\mathbf{c}(\bar{t}):=$ c(t($\bar{t}$)), $\bar{t}_i \leq \bar{t} \leq \bar{t}_f$ is a reparameterization of c, with $\frac{dt}{d\bar{t}} > 0$, then L(c) = L($\mathbf{c}$), and hence the length of a curve is well-defined.

Finsler Geometry generalizes this notion of length using a function f:TM\{0}→$\mathbb{R}^+$, which is such that $\forall \lambda \in \mathbb{R}^+$ and v∈TM\{0}, $f(\lambda v) = \lambda^2 f(v)$. We can then use f to define the length of curves c=c(t), a≤t≤b, in M by

$$L_f(c) := \int_a^b |f(c')|^{1/2} dt .$$

(We do not really need the absolute value signs in the above integrand, but they are required when we consider pseudo-Finsler Spaces in which f maps into $\mathbb{R}$.) Since f is positively homogeneous of degree 2, $L_f(c)$ is well-defined; i.e., independent of parameterization.

If f is to define a Finsler Space, then there exists a second very important condition that it must satisfy. But in order to not be too repetitive I shall hold off stating that condition until we get to Pseudo-Finsler Spaces in just a moment.

If $V_n$=(M,g) is a Riemannian or pseudo-Riemannian space, then g naturally gives rise to a function f on TM defined by

$$f(v) = |g(v,v)|.$$

We shall call Finsler spaces of this type, trivial Finsler Spaces. Examples of non-trivial Finsler spaces are provided in [2] and [3].

I would like to point out that the Finsler function is usually denoted by F in the literature on the subject. However, physicists usually use F for the electromagnetic field tensor. Thus to avoid confusion, and to also allow the introduction of electromagnetic fields into our discussion, I chose f to denote the Finsler function.

I shall now state the formal definition of pseudo-Finsler Spaces following the presentation given by Bejancu & Farran [4].

Let M be an n-dimensional smooth manifold with tangent bundle TM, and let $\pi$: TM$\rightarrow$M denote the natural projection. So if v$\in$TM is a tangent vector at P$\in$M, then $\pi$(v)=P. If x is a chart of M with domain U, it naturally gives rise to a chart $(\chi,y)$ in TM, with domain $\pi^{-1}$U, which is defined by

$$v = v^i \frac{\partial}{\partial x^i}\bigg|_{\pi(v)} \in \pi^{-1}U \rightarrow (\chi(v),\, y(v)),$$

with $\chi:=x\circ\pi$ , $y(v):=[v^i]$, and Einstein's summation convention is being used. If x' is another chart of M with domain U', and U$\cap$U'$\neq$ø, then on U$\cap$U' we can write $x'^i=x'^i(x^i)$, and conversely $x^i=x^i(x'^j)$. In this case the standard charts $(\chi',y')$ and $(\chi,y)$ are overlapping charts of TM. On the overlap we have

$$\chi'^j = \chi'^j(\chi^i) \ , \ y'^j = y^h \frac{\partial x'^j}{\partial x^h} \circ \pi \qquad\qquad\qquad \text{Eq.1.5a}$$

and conversely

$$\chi^j = \chi^j(\chi^{'i}) \ , \quad y^j = y^{'h}\frac{\partial x^j}{\partial x^{'h}} \circ \pi \qquad\qquad\qquad \text{Eq.1.5b}$$

Now for the definition of a pseudo-Finsler space. Consider a positive integer q<n, and a smooth function f: N→ℝ, where N is an open submanifold of TM with π(N) = M, and θ(M)∩N=ø, where θ:M→TM is the zero section. N is required to be invariant under dilation: *i.e.*, ∀ λ∈ℝ⁺, and v∈N, λv∈N. Lastly f is required to satisfy the following two conditions:

(i) f is positively homogeneous of degree 2, *i.e.,* ∀ λ∈ℝ⁺ and v∈N, f(λv) = λ²f(v); and

(ii) if v ∈N and (χ,y) is a standard chart of TM at v, then the matrix

$$\left[ \frac{\frac{1}{2}\partial^2 f(v)}{\partial y^i \partial y^j} \right]$$

defines a quadratic form on ℝⁿ with q negative eigenvalues and n−q positive eigenvalues, 0<q<n.

We say that the triple $F^n$:= (M,N,f) is a pseudo-Finsler Space of index q, and refer to f as the pseudo-Finsler function.  When q=1 or n−1, $F^n$ is said to be a Lorentzian Finsler Space.

One should note that due to Eqs.1.5a and 1.5b the signature of f is well-defined. For suppose that (χ',y') is another standard chart at v.  Then we have

$$\left.\frac{\partial f}{\partial y^{'i}}\right|_v = \left.\frac{\partial f}{\partial \chi^j}\frac{\partial \chi^j}{\partial y^{'i}}\right|_v + \left.\frac{\partial f}{\partial y^j}\frac{\partial y^j}{\partial y^{'i}}\right|_v \qquad\qquad \text{Eq.1.6}$$

Since $\chi^j = \chi^j(\chi^{'k})$, the first term on the right-hand side of Eq.1.6 vanishes, and we can use Eq.1.5b to rewrite Eq.1.6 as

$$\frac{\partial f}{\partial y^{ti}}\bigg|_v = \frac{\partial f}{\partial y^j}(v)\,\frac{\partial x^j}{\partial x^{ti}}(\pi(v))\ .$$

It should now be clear that

$$\frac{\partial^2 f}{\partial y^{ti}\partial y^{tj}}(v) = \frac{\partial^2 f}{\partial y^h\partial y^k}(v)\,\frac{\partial x^h}{\partial x^{ti}}\,\frac{\partial x^k}{\partial x^{tj}}(\pi(v)) \qquad . \qquad\qquad\qquad \text{Eq.1.7}$$

Hence the index of $\dfrac{\partial^2 f(v)}{\partial y^i\partial y^j}$ is independent of the standard chart of TM that we use. From Eq.1.8 it should be evident how the higher order "vertical derivatives" of f would transform under coordinate transformations, where by vertical derivatives I mean derivatives with respect to the $y^i$'s, which are the coordinates on the (vertical) fibres of TM.

Eq.1.8 also shows us that if we have a vector field $\xi$ on M which is such that $\xi_p \in N \ \forall\ P \in M$, then we can define a pseudo-Riemannian metric tensor $g_\xi$ on M using f and $\xi$. To do this let x be any chart of M, with corresponding standard chart $(\chi,y)$ of TM. Then we define the x components of $g_\xi$ by

$$g_{\xi,ij} := g_\xi\!\left(\frac{\partial}{\partial x^i}, \frac{\partial}{\partial x^j}\right) := \tfrac{1}{2}\frac{\partial^2 f}{\partial y^i\partial y^j}(\xi) \qquad . \qquad\qquad\qquad \text{Eq.1.8}$$

Due to Eq.1.7, $g_\xi$ is a well-defined pseudo-Riemannian metric tensor on M. Similarly the $k^{th}$ order vertical derivatives of f (k>0) could be used in conjunction with the vector field $\xi$ to construct symmetric (0,k) tensor fields on M.

In summary, if we have a pseudo-Finsler Space $F^n = (M,N,f)$, of index q, and a vector field $\xi$ on M whose range lies in N, then we also have a pseudo-Riemannian Space $V_n = (M,g_\xi)$ of index q.

Now let us confine our attention to n=4, with index 1, so we have a Lorentzian Finsler Space. Earlier in this section I showed how we could introduce energy scalar fields $\mathscr{E}$ and $\mathscr{E}_T$ on an open submanifold of TM (*see*, Eqs.1.2 and 1.3). So the natural question to ask is: Can we use an energy scalar field and f as the basis of a gravitational field theory? We could then use f and $\xi$ to give us a Lorentzian metric tensor on M in the manner described above. The scalar fields $\mathscr{E}$ and f (or $\mathscr{E}_T$ and f) in TM could be thought of as a generalization of the variables appearing in Poisson's classical gravitational equation. To complete our task all we need are some field equations on TM governing a pair of scalar fields. Unfortunately no such equations exist due to the following

**Proposition:** If $\mathscr{L}$ is a Lagrange scalar density on an n-dimensional manifold (n>1) which is a concomitant of two scalar fields, $\psi_1$ and $\psi_2$, along with their derivatives of arbitrary order, then

$$\frac{\delta\mathscr{L}}{\delta\psi_1} \equiv 0 \text{ and } \frac{\delta\mathscr{L}}{\delta\psi_2} \equiv 0 .$$

**Proof:** Since $\mathscr{L}$ is a scalar density, its associated Euler-Lagrange tensors are interrelated by a Noether type mathematical identity. In [5] (*see,* the bottom of page 49 in [5]) I derived that identity for the case in which the Lagrangian is a concomitant of a metric tensor field, a covariant vector field and a scalar field, along with their derivatives. That identity is easily generalized to the present case, and is given by

$$\frac{\delta \mathcal{L}}{\delta \psi_1} \, d\psi_1 + \frac{\delta \mathcal{L}}{\delta \psi_2} \, d\psi_2 \; \equiv 0 \; . \qquad\qquad \text{Eq.1.9}$$

Eq.1.9 must hold $\forall$ pair of scalar fields. Since n>1 this is only possible when

$$\frac{\delta \mathcal{L}}{\delta \psi_1} = \frac{\delta \mathcal{L}}{\delta \psi_2} \equiv 0. \; \blacksquare$$

This proposition destroys our aspiration to construct gravitational field equations governing an energy scalar field and f on TM using a Lorentzian Finsler Space and a vector field $\xi$ on M. However, my real desire is to construct some sort of generalized scalar-tensor theory using a Finsler-like structure. This task will be tackled in the following sections.

## Section 2: Cofinsler Spaces and Scalar-Scalar Field Theories

Suppose that we have a 4-dimensional Lorentzian Finsler Space $F^4 = (M,N,f)$ and a scalar field $\varphi$ on M. Is there a simple way for us to use f and $\varphi$ to endow M with a Lorentzian metric tensor? At first you might think that if the gradient of $\varphi$, $\nabla\varphi$, lies in N then we could use it to build a metric on M. But to build $\nabla\varphi$ you need a metric tensor, g, on M, since locally $\nabla^i\varphi = g^{ij}\varphi_{,j}$. Unfortunately there is no easy way to use $F^4$ and $\varphi$ to endow M with a Lorentzian structure. Thus we shall have to take a different approach if we want to use the scalar field φ in the construction of a Lorentzian metric tensor on M. To that end we note that dφ naturally provides us with a map from M into T*M, the cotangent bundle of M. Hence what we need is a smooth map from T*M into $\mathbb{R}$, whose second derivatives in the vertical direction (*i.e.,* in the direction tangent to the fibres of T*M) constitute a Lorentzian quadratic form when evaluated at dφ. I shall now make these notions a bit more precise.

Let M be an n-dimensional manifold with cotangent bundle T*M. Recall that $T^*M := \bigcup_{P \in M} T_P^*M$, where $T_P^*M$ is the dual space of $T_PM$, the tangent space to M at P. We let $\pi : T^*M \to M$ denote the canonical (or natural) projection, which maps a covector $\omega \in T_P^*M$ to P: $\pi(\omega) = P$. If x is a chart of M with domain U then it naturally gives rise to a standard chart $(\chi, y)$ of T*M with domain $\pi^{-1}U$ defined by $\chi := x \circ \pi$ and $y(\omega) = y(\omega_i dx^i|_p) := (\omega_1, \ldots, \omega_n) \equiv (\omega_i)$. (There should be no confusion between the standard charts of TM and T*M, since it should always clear what space we are working in.) So if we write $y : \pi^{-1}U \to \mathbb{R}^n$, as $y = (y_1, \ldots, y_n) \equiv (y_i)$, then $y_i(\omega) = \omega_i$, where $\omega = \omega_i dx^i|_p$. Now suppose x and x' are overlapping charts of M with domains U and U'. Then $(\chi, y)$ and $(\chi', y')$ are overlapping charts in T*M. On the overlap we have

$$\chi^i = \chi^i(\chi'^j), \quad \chi'^i = \chi'^i(\chi^j) \qquad \text{Eq.2.1a}$$

and

$$y_i = y'_j \frac{\partial \chi'^j}{\partial \chi^i}, \quad y'_i = y_j \frac{\partial \chi^j}{\partial \chi'^i}, \qquad \text{Eq.2.1b}$$

where the definition of partial differentiation on a manifold was used to make

the substitutions

$$\frac{\partial \chi^j}{\partial \chi^{\prime i}} = \frac{\partial x^j}{\partial x^{\prime i}} \circ \pi \quad \text{and} \quad \frac{\partial \chi^{\prime j}}{\partial \chi^i} = \frac{\partial x^{\prime j}}{\partial x^i} \circ \pi \quad . \qquad\qquad \text{Eq.2.2}$$

With these preliminaries disposed of I can now define an n-dimensional Pseudo-Cofinsler Space. Let f: T*M→ℝ be a smooth function defined on an open submanifold N of T*M where πN = M. Choose a positive integer q such that q<n. If x is a chart of M with domain U, and corresponding standard chart (χ,y) of T*M with domain $\pi^{-1}U$, then we require that $\forall\ \omega \in \pi^{-1}U \cap N$ the matrix

$$\left[ \tfrac{1}{2} \frac{\partial^2 f}{\partial y_i \partial y_j} \right]\Big|_\omega \qquad\qquad \text{E q.2.3}$$

defines a quadratic form on ℝ$^n$ of index q. Due to Eqs.2.1a and 2.1b the index of the matrix presented in Eq.2.3 is independent of the standard chart chosen at $\omega \in$T*M. When these conditions are met the triple *CF*$^n$:=(M,N,f) is called an n-dimensional Pseudo-Cofinsler Space of index q, and f is called the Cofinsler function. When q = 1 or n−1, *CF*$^n$ is called a Lorentzian Cofinsler Space. In what follows our Lorentzian Cofinsler Spaces will have index 1.

Note that in the definition of Cofinsler Spaces I have not placed any homogeneity constraints upon f, as is done in the Finsler case. This was done because there is no natural way to "lift" curves in M into T*M to compute their lengths using f. Consequently there is no reason to require f to be homogeneous, nor N to be dilation invariant.  In addition, N∩θ(M), need not be empty, where θ:M→T*M is the zero section.

Suppose we have an n-dimensional pseudo-Cofinsler Space $CF^n = (M,N,f)$ and a scalar function φ on M. If the range of dφ lies in N; *i.e.,* dφ(M)⊂N, then we can define the contravariant components of a pseudo-Riemmanian metric tensor $g_φ$ on M by

$$g_φ^{ij} := g_φ(dx^i,dx^j) := \tfrac{1}{2}\frac{\partial^2 f}{\partial y_i \partial y_j}(dφ) \,, \qquad\qquad \text{Eq.2.4}$$

where x is any chart of M with corresponding standard chart (χ,y) for T*M.  Due to Eqs.2.1 and 2.2 it is clear that the $g_φ^{ij}$'s are the local contravariant components of a pseudo-Riemannian metric tensor on M.

I shall now present an example of a 4-dimensional Lorentzian Cofinsler Space that will play a significant roll in what follows.  To see where this example comes from recall that the FLRW (:=Friedmann-Lemaître-Roberston -Walker) metric is given by

$$ds^2 = -dt^2 + a(t)^2\left(\frac{dr^2}{1-kr^2} + r^2 d\theta^2 + r^2\sin^2\theta d\varphi^2\right) \,, \qquad\qquad \text{Eq.2.5}$$

where k is a constant with units of $(length)^{-2}$, r has units of length, a(t) is unitless and c=1.  k determines the curvature of the spaces t = constant, which are 3-surfaces of constant curvature. To make matters even simpler I shall take k=0 in Eq.2.5, and so our line element becomes:

$$ds^2 = -dt^2 + a(t)^2(du^2 + dv^2 + dw^2) ,\qquad\qquad\qquad Eq.2.6$$

where $x \equiv (x^0, x^1, x^2, x^3) := (t,u,v,w)$, are the standard coordinates of $\mathbb{R}^4$, and the 3-spaces

$t$ = constant are flat. What we would like is a function f in $T^*\mathbb{R}^4$ which is such that

when we construct its associated metric g using Eq.2.4, and a scalar field $\varphi$ on $\mathbb{R}^4$, we

get a line element similar to the one presented in Eq.2.6. Well, sure, we could always

choose the trivial function $f_T$ defined by

$$f_T := -y_t^2 + a(t\circ\pi)^{-2}((y_u)^2 + (y_v)^2 + (y_w)^2),$$

where $((t,u,v,w)\circ\pi, (y_t,y_u,y_v,y_w))$ is the standard chart of $T^*\mathbb{R}^4$ determined by the chart

$(t,u,v,w)$ of $\mathbb{R}^4$. (When working with $T^*\mathbb{R}^4$ I shall adopt the custom of dropping the

projection function $\pi$ from the standard chart, since $T^*\mathbb{R}^4$ is obviously diffeomorphic to $\mathbb{R}^4 \times \mathbb{R}^4$.) For this choice of Cofinsler function $d\varphi$ would actally play no roll in the

associated metric tensor. A more useful choice of Cofinsler function is

$$f := -y_t^2 + y_t^{-2}((y_u)^2 + (y_v)^2 + (y_w)^2) . \qquad\qquad\qquad Eq.2.7$$

When $\varphi = \varphi(t)$, this Cofinsler function gives us

$$[g_\varphi^{\ ij}] = \left[ \tfrac{1}{2}\frac{\partial^2 f}{\partial y^i \partial y^j}(d\varphi) \right] = diag(-1,\ \varphi'^{-2},\ \varphi'^{-2},\ \varphi'^{-2}) \qquad\qquad Eq.2.8$$

where a prime denotes a derivative with respect to t. Evidently $g_{ij}dx^i dx^j$ (where here,

and in what follows, I shall drop the subscipt $\varphi$ on the covariant and contravariant form of the metric tensor) gives us the line element presented in Eq.2.6 with $a(t) = \varphi'(t)$.

At this point some remarks concerning units is in order. In what follows we shall use geometrized units in terms of which c=G=1, and all dimensioned quantities will have units of length, $\ell$, to some power. To convey the fact that a physical quantity $\Omega$ has units of $\ell^\lambda$ we shall write $\Omega \sim \ell^\lambda$. In terms of these units the coordinates t, u, v, w $\sim \ell^1$, and $g_{ij} \sim \ell^0$. Consequently $\varphi' \sim \ell^0$, and hence for us $\varphi \sim \ell^1$. In terms of geometrized coordinates the $y_i$ (as well as the $y^i$ in TM) are unitless. This follows from the fact that

$$y_i(d\varphi) = \frac{\partial \varphi}{\partial x^i} \sim \ell^0.$$

Since $g^{ij} \sim \ell^0$, we can now use Eq.2.4 to deduce that $f \sim \ell^0$, as it does in Eq.2.7.

From Eq.2.6 we know that the metric on the t=constant slices is

$$(\varphi')^2(du^2 + dv^2 + dw^2) \ .$$

Thus the unitless quantity $|\varphi'|$ plays the roll of a scale factor in these slices. To get the physical distance between any two distinct points in a slice you just multiply their Euclidean distance by $|\varphi'|$. An interesting thing happens when $\varphi' \rightarrow 0^+$. When that occurs the metric is telling us that the physical distance between any two points in a slice is vanishing, while topologically the slice still remains $\mathbb{R}^3$. So when $\varphi'=0$ our physical spacetime is not crushed out of existence but remains intact. More will be said about this later when we deal with physical interpretation of solutions to the various field equations which I shall construct.

The largest subset N of $T^*\mathbb{R}^4$ upon which the function f presented in Eq.2.7 is defined and smooth is N = $\{((t,u,v,w),(y_t,y_u,y_v,y_w))|\ y_t \neq 0\}$. It should be noted that f is not homogeneous on N.

To determine if f really is a Cofinsler function on $N \subset T^*\mathbb{R}^4$, we need to examine the matrix $g^{ij}$ on N and not just for $d\varphi$. It is easy to use Eq.2.7 to show that

$$[g^{ij}] = \begin{bmatrix} -1 & -2y_t^{-3}y_u & -2y_t^{-3}y_v & -2y_t^{-3}y_w \\ -2y_t^{-3}y_u & y_t^{-2} & 0 & 0 \\ -2y_t^{-3}y_v & 0 & y_t^{-2} & 0 \\ -2y_t^{-3}y_w & 0 & 0 & y_t^{-2} \end{bmatrix} \qquad \text{Eq.2.9}$$

and, with some effort, its inverse $[g_{ij}]$ is found to be

$$[g_{ij}] =$$

$$D^{-1}\begin{bmatrix} -1 & -2y_t^{-1}y_u & -2y_t^{-1}y_v & -2y_t^{-1}y_w \\ -2y_t^{-1}y_u & [y_t^2+4y_t^{-2}(y_v)^2+4y_t^{-2}(y_w)^2] & -4y_t^{-2}y_uy_v & -4y_t^{-2}y_uy_w \\ -2y_t^{-1}y_v & -4y_t^{-2}y_uy_v & [y_t^2+4y_t^{-2}(y_u)^2+4y_t^{-2}(y_w)^2] & -4y_t^{-2}y_vy_w \\ -2y_t^{-1}y_w & -4y_t^{-2}y_uy_w & -4y_t^{-2}y_vy_w & [y_t^2+4y_t^{-2}(y_u)^2+4y_t^{-2}(y_v)^2] \end{bmatrix}$$
$$\text{Eq.2.10}$$

where

$$D := 1 + 4y_t^{-4}(y_u^2 + y_v^2 + y_w^2) \qquad \text{Eq.2.11}$$

and

$$\det [g_{ij}] = -y_t^6 D^{-1} . \qquad \text{Eq.2.12}$$

In view of Eq.2.12 it is clear that f defines a Lorentzian Cofinsler function on $N \subset T^*\mathbb{R}^4$.

In passing I would like to point out that a Finsler or Pseudo-Finsler Space of the form $(\mathbb{R}^n, N, f)$ which is such that the Finsler function f is only a function of the $y^i$ coordinates is called a Minkowski Finsler or Minkowski Pseudo-Finsler Space. In keeping with that nomenclature we probably should refer to the Cofinsler Space just introduced as a Minkowskian Lorentzian Cofinsler Space. However, I shall not do that since, not only is it a mouthful of terminology, but Minkowski Spaces have a different meaning for Relativists, and I do not want to cause any undue confusion.

I shall call a Lorentzian Spacetime, $V_4$=(M,g) endowed with a scalar field φ, in which the metric tensor g arises from a Lorentzian Cofinsler Space $CF^4$=(M,N,f) in the manner presented in Eq.2.4, as a scalar-scalar theory. It is tempting to think of f as a generating function, or as a potential for the Lorentzian metric tensor on M. But that is not quite right, since both f and φ are required to generate g. Our next task is to develop field equations for scalar-scalar theories. This will be done forthwith.

## Section 3: Lagrangians and Field Equations

Suppose that we have a scalar-scalar field theory based on the Lorentzian Cofinsler Space $CF^4$=(M,N,f) and the scalar field φ. We would like to build a Lagrangian from f and φ. The first problem we encounter is that f and φ are defined on different spaces. Well, this is easily circumvented by letting Φ:= φ∘π, where π:T*M→M is the natural projection. Now f and Φ are two smooth scalar fields on T*M. Unfortunately it is impossible to use these scalar fields on T*M to build a useful Lagrangian due to the **Proposition** presented at the end of **Section 1**.

So let us now consider the possibility of deriving field equations on M using f and φ. To that end let $\mathcal{L}$ be any second-order Lagrange scalar density on M which is a concomitant of a metric tensor and scalar field φ. We replace the metric tensor by our cofinsler metric tensor built from the vertical derivatives of f evaluated at dφ, to obtain a Lagrangian of the form

$$\mathcal{L} = \mathcal{L}(g^{ij}(d\varphi); g^{ij}(d\varphi)_{,h}; g^{ij}(d\varphi)_{,hk}; \varphi; \varphi_{,h}; \varphi_{,hk})  \qquad \text{Eq.3.1}$$

which is locally well-defined on the coordinate domains of M. However, we are now confronted with a problem which is unique to Cofinsler (and Finsler Theory for that matter); *viz*., how does one go about computing the variational derivatives of a Lagrangian of the form presented in Eq.3.1? Say we want to vary φ holding f in T*M fixed. Since the $g^{ij}$'s are functions of dφ, any variation of φ will lead to a variation of the $g^{ij}$'s. Conversely, the general variation of

$$g^{ij} = \tfrac{1}{2} \frac{\partial f}{\partial y_i \partial y_j} \bigg|_{d\varphi}$$

would require a variation of both f and φ. In my experience with classical field theories I have not encountered such a convoluted variational problem. Usually when one encounters a Lagrangian which is a concomitant of various fields (all defined on the same manifold), one computes the variational derivative by holding all fields fixed except for one which is varied. That is not possible with the Lagrangian given in Eq.3.1. I shall now describe a method that we can use to circumvent this impediment to computing field equations for scalar-scalar theories. To that end we need to re-examine some very familiar equations from General Relativity.

Let us consider the derivation of the Schwarschild solution. Recall that in this situation we are looking for a static, spherically symetric, asymptotically flat solution to Einstein's equations $G^{ij} = 0$. The symmetry demands imply that we can choose a coordinate system so that our ansatz metric has the form

$$ds^2 = -e^{\nu}dt^2 + e^{\lambda}dr^2 + r^2 d\theta^2 + r^2\sin^2\theta\, d\varphi^2 \qquad\qquad \text{Eq.3.2}$$

where $\nu$ and $\lambda$ are functions of r with $\theta$ and $\varphi$ being spherical polar coordinates. Conventionally one plugs this metric into $G^{ij} = 0$ to obtain the following equations (*c.f.* Adler, Bazin & Schiffer [6])

$$R_{00} = \tfrac{1}{2}e^{(\nu-\lambda)}[\nu'' + \tfrac{1}{2}(\nu')^2 - \tfrac{1}{2}\nu'\lambda' + 2\nu'r^{-1}] = 0,$$

$$R_{11} = \tfrac{1}{2}[\nu'' + \tfrac{1}{2}(\nu')^2 - \tfrac{1}{2}\nu'\lambda' - 2\lambda'r^{-1}] = 0,$$

$$R_{22} = e^{-\lambda}[1 + \tfrac{1}{2}r(\nu' - \lambda')] - 1 = 0,$$

$$R_{33} = \sin^2\theta\, R_{22}\,,$$

where a prime denotes a derivative with respect to r, and the other $R_{ij}$ vanish. These equations are readily solved and yield $\lambda = -\nu$ with $e^{\nu} = 1 - 2mr^{-1}$, in terms of units with c=G=1.

However, there exist a second way to proceed to get the equations for $\lambda$ and $\nu$, which I shall describe, even though it is a bit impractical. We know that the Einstein field equations $G^{ij}=0$, can be derived by varying the metric tensor in $\mathcal{L}_E :=(-g)^{\frac{1}{2}}R$. So let us build $\mathcal{L}_E$ from the line element given in Eq.3.2. This yields

$$\mathcal{L}_E = -2\,\sin\theta\;e^{\frac{1}{2}(\nu-\lambda)}(r\lambda' - 1 + e^\lambda)\;.$$

If we vary the scalar fields $\lambda$ and $\nu$ in this Lagrangian we obtain the following Euler-Lagrange equations:

$$\frac{\delta\mathcal{L}_E}{\delta\lambda} = \sin\theta\;e^{\frac{1}{2}(\nu-\lambda)}(1 + r\nu' - e^\lambda) = 0$$

and

$$\frac{\delta\mathcal{L}_E}{\delta\nu} = \sin\theta\;e^{\frac{1}{2}(\nu-\lambda)}(1 - r\lambda' - e^\lambda) = 0\;.$$

Using the assumption of asymptotic flatness, we can solve this pair of equations to obtain the usual Schwarzschild result.

This observation concerning the Schwarzschild equations suggests an approach to deriving field equations from a Lagrangian of the form presented in Eq.3.1. We should begin by guessing a form of f suitable to the problem at hand. Plug that ansatz f into $\mathcal{L}$ along with $\varphi$, and then vary the fields in $\mathcal{L}$, just as we did in $\mathcal{L}_E$ above. The Lagrangian that we get in this manner will be called a hidden scalar Lagrangian and will be a concomitant of $\varphi$, its derivatives and perhaps the local coordinates of M (if f locally depends upon the $\chi^i$ coordinates as well as the $y_i$, as it will for those f's that generate a FLRW metrics with $k \neq 0$).

To illustrate this approach to obtaining field equations for scalar-scalar field theories I shall derive some cosmological solutions for various choices of $\mathcal{L}$. For these solutions it will be assumed that f has been chosen as in Eq.2.7, with $\varphi=\varphi(t)$ and $g^{ij}$ given by Eq.2.8. I have selected this form of f since it generates a FLRW line element, and I shall refer to it as the FLRW Cofinsler function.

Let us begin by examining the form of the Horndeski Lagrangians $L_H$ (*see,* [5] or [7]) when

$$[g_{ij}] = \text{diag}(-1, (\varphi')^2, (\varphi')^2, (\varphi')^2) \,. \qquad \text{Eq.3.3}$$

In general the scalar-tensor Lagrangian $L_H$ is given by

$$L_H = L_2 + L_3 + L_4 + L_5 \qquad \text{Eq.3.4}$$

where

$$L_2 := g^{\frac{1}{2}} K(\varphi, X) \qquad \text{Eq.3.5}$$

$$L_3 := g^{\frac{1}{2}} G_3(\varphi, X) \,\Box\varphi \qquad \text{Eq.3.6}$$

$$L_4 := g^{\frac{1}{2}} G_4(\varphi, X) \, R - 2g^{\frac{1}{2}} G_{4,X}(\varphi, X)((\Box\varphi)^2 - \varphi^{ab}\varphi_{ab}) \qquad \text{Eq.3.7}$$

and

$$L_5 := g^{\frac{1}{2}} G_5(\varphi, X) \, \varphi_{ab} G^{ab} + \tfrac{1}{3} g^{\frac{1}{2}} G_{5,X}(\varphi, X)((\Box\varphi)^3 - 3\Box\varphi(\varphi^{ab}\varphi_{ab}) + 2\varphi^a_{\ b}\varphi^{bc}\varphi_{ac}), \ \text{Eq.3.8}$$

where $g := |\det g_{ab}|$, $X := g^{ab}\varphi_{,a}\varphi_{,b}$, "$_{,X}$" denotes a partial derivative with respect to X, "$_{,a}$" denotes a partial derivative with respect to the local coordinate $x^a$, $\varphi_{ab}$ denotes the components of the second covariant derivative of $\varphi$, $\Box\varphi := g^{ab}\varphi_{ab}$, $R := g^{ij}R_{ij}$, $R_{ij} := R_i{}^h{}_{jh}$ $G^{ij} := R^{ij} - \tfrac{1}{2}g^{ij}R$, and $R_i{}^h{}_{jk} := \Gamma^h{}_{ij,k} - \Gamma^h{}_{ik,j} + \Gamma^m{}_{ij}\Gamma^h{}_{mk} - \Gamma^m{}_{ik}\Gamma^h{}_{mj}$ with $\Gamma^h{}_{ij} := \tfrac{1}{2}g^{hm}(g_{im,j} + g_{jm,i} + -g_{ij,m})$.

The form of $L_H$ given above differs from what I present in [5] and [7], and was introduced by Deffayet, *et al.,* in [8]. Kobayashi, *et al.,* [9], then showed that the two expressions for $L_H$ were in fact identical, up to the addition of divergences.

Using the metric tensor given in Eq.3.3 we find that the nonzero components of $\Gamma^h_{ij}$, $R_{ij}$, $R$, $G_{ij}$ and $\varphi_{ij}$ are given by

$$\Gamma^t_{uu} = \Gamma^t_{vv} = \Gamma^t_{ww} = \varphi'\varphi'' \qquad\qquad Eq.3.9$$

$$\Gamma^u_{tu} = \Gamma^v_{tv} = \Gamma^w_{tw} = \varphi''/\varphi' \qquad\qquad Eq.3.10$$

$$R_{tt} = -3\varphi'''/\varphi' \; , \; R_{uu} = R_{vv} = R_{ww} = \varphi'\varphi''' + 2(\varphi'')^2 \; , \; R = 6(\varphi'''/\varphi' + (\varphi''/\varphi')^2) \qquad Eq.3.11$$

$$G_{tt} = 3(\varphi''/\varphi')^2 \; , \; G_{uu} = G_{vv} = G_{ww} = -(2\varphi'''\varphi' + (\varphi'')^2) \qquad\qquad Eq.3.12$$

and

$$[\varphi_{ab}] = diag(\varphi_{tt}, \varphi_{uu}, \varphi_{vv}, \varphi_{ww}) = diag(\varphi'', -\varphi''(\varphi')^2, -\varphi''(\varphi')^2, -\varphi''(\varphi')^2) \qquad Eq.3.13$$

with

$$\square\varphi = -4\varphi'' , \; X = -(\varphi')^2 \; , \; g^{\frac{1}{2}} = (\varphi')^3 \qquad\qquad Eq.3.14$$

where a prime denotes a derivative with respect to t, repeated indices t, u, v and w are not summed over, and it is assumed that $\varphi'>0$. Employing Eqs. 3.10–3.14 we find that the Horndeski Lagrangians given in Eqs.3.5–3.8 become

$$L_2 = (\varphi')^3 K(\varphi, -(\varphi')^2) \qquad\qquad Eq.3.15$$

$$L_3 = -4G_3(\varphi, -(\varphi')^2)(\varphi')^3\varphi'' \qquad\qquad Eq.3.16$$

$$L_4 = 6G_4(\varphi, -(\varphi')^2)[(\varphi')^2\varphi''' + \varphi'(\varphi'')^2] - 24G_{4,X}(\varphi, -(\varphi')^2)(\varphi')^3(\varphi'')^2 \qquad Eq.3.17$$

$$L_5 = 6\varphi'G_5(\varphi, -(\varphi')^2)[(\varphi'')^3 + \varphi'\varphi''\varphi'''] - 8G_{5,X}(\varphi, -(\varphi')^2)(\varphi')^3(\varphi'')^3 \; . \qquad Eq.3.18$$

At first glance the Lagrangians $L_4$ and $L_5$ seem to be of third order in $\varphi$, but the third order terms can be assimilated into divergences. Upon doing this and dropping the divergences, $L_4$ and $L_5$ become $L_{4SO}$ and $L_{5SO}$ ("SO" meaning "second-order"), and are given by

$$L_{4SO} = -6G_4(\varphi, -(\varphi')^2)\varphi'(\varphi'')^2 - 6G_{4,\varphi}(\varphi, -(\varphi')^2)(\varphi')^3\varphi'' +$$

$$-12G_{4,X}(\varphi, -(\varphi')^2)(\varphi')^3(\varphi'')^2 \qquad \text{Eq.3.19}$$

and

$$L_{5SO} = -3G_{5,\varphi}(\varphi, -(\varphi')^2)(\varphi')^3(\varphi'')^2 - 2G_{5,X}(\varphi, -(\varphi')^2)(\varphi')^3(\varphi'')^3 . \qquad \text{Eq.3.20}$$

At this point all savvy readers have probably leaped to their feet screaming and pointing at $L_{4SO}$ and $L_{5SO}$ as non-degenerate second-order Lagrangians, rotten with Ostrogradsky [10] type singularities (*cf.,* Woodward [11]), and as such they should be thrown to the dogs! But I say, let us not be too hasty. One of the problems associated with Lagrangians like $L_{4SO}$ and $L_{5SO}$ is that they can lead to multiple vacuum states. This would be a good thing if we were looking for an equation that predicted the multiverse, since a multiverse would need a separate vacuum for each of the individual universes. So, stay tuned!

In view of what we saw when we examined the Schwarzschild solution of Einstein's equations, one might think that if we vary $\varphi$ in the Lagrangians presented in Eqs.3.15, 3.16, 3.19 and 3.20, the equations that we obtain will somehow just be what one gets from the usual scalar-tensor equations derived from the Lagrangians $L_2$, . . . , $L_5$ when the metric has the form presented in Eq.3.4. However, that is not the case. For example let us consider the simple Lagrangian

$$L_S := g^{\frac{1}{2}} K(\varphi) . \qquad \text{Eq.3.21}$$

The equations $\dfrac{\delta L_S}{\delta g_{ij}} = 0$, and $\dfrac{\delta L_S}{\delta \varphi} = 0$ are

$g^{\frac{1}{2}} g^{ij} K = 0$ and $g^{\frac{1}{2}} K_{,\varphi} = 0$ ,

where "$_{,\varphi}$" denotes a derivative with respect to $\varphi$. Consequently, K must vanish. But

if we evaluate $L_S$ for the metric given in Eq.3.4 we obtain

$$L_S = (\varphi')^3 K . \qquad\qquad Eq.3.22$$

For this form of $L_S$ the equation $\dfrac{\delta L_S}{\delta \varphi} = 0$ is given by

$$K_{,\varphi}(\varphi')^3 + 3K\varphi'\varphi'' = 0,$$

which implies (since $\varphi' \neq 0$) that

$$K(\varphi')^3 = \beta^3 , \qquad\qquad Eq.3.23$$

where $\beta$ is a constant. This equation can be integrated to yield

$$\beta t + \kappa = \int K^{1/3} d\varphi, \qquad\qquad Eq.3.24$$

with $\kappa$ being an integration constant. We can now get numerous expressions for $\varphi$ by

using different choices of K. *E.g.,* if we want the Universe to grow exponentially as

t increases from 0, we could choose $K = \varphi^{-3}$ in Eq.3.24 to get

$$\varphi = \alpha\, e^{\beta t}, \text{ and } \varphi' = \alpha\beta\, e^{\beta t} , \qquad\qquad Eq.3.25$$

where $\alpha \sim \ell^1$, $\beta \sim \ell^{-1}$, and $\alpha\beta > 0$.

If we choose $K = \varphi^{3s}$, $s \neq -1$, then we can use Eq.3.24 to find that

$$\varphi = [(s+1)(\beta t + \kappa)]^{1/(s+1)} \text{ and } \varphi' = \beta[(s+1)(\beta t + \kappa)]^{-s/(s+1)} . \qquad Eq.3.26$$

So if we take $\kappa = 0$, and choose s so that $-\frac{1}{2} < s < 0$, then we obtain a Universe that expands explosively from t=0, where $\varphi'$ has a vertical tangent. If we take $s = -2/5$ then the metric tensor in this model is identical to the one appearing in a non-empty cosmological model of Einstein's theory (*see,* page 364 of Adler, *et al.,* [6]).

Thus we see that the simple Lagrangian, $L_S$, given in Eqs.3.21 and 3.22 is capable of providing us with interesting cosmological models. What is the Hamiltonian for this Lagrangian? Since $L_S$ is of first order in $\varphi$ we have

$$H_S = P_S \varphi' - L_S ,$$
where
$$P_S := \frac{\partial L_S}{\partial \varphi'} = 3(\varphi')^2 K .$$

Consequently we have $H_S = 2(\varphi')^3 K$, and hence due to Eq.3.23 we see that when the field equations are satisfied

$$H_S = 2\beta^3 ,$$

a constant, which vanishes only for the trivial solution.

Let us now turn our attention to the construction of solutions to the equations derived from the other Lagrangians appearing in $L_H$.

For a Lagrangian, L, which is third-order in $\varphi = \varphi(t)$, the Ostrogradsky Hamiltonian is given by (*see,* Woodard [11])

$$H := P_1 \varphi' + P_2 \varphi'' + P_3 \varphi''' - L, \qquad\qquad Eq.3.27$$

where
$$P_1 := \frac{\partial L}{\partial \varphi'} - \frac{d}{dt}\frac{\partial L}{\partial \varphi''} + \frac{d^2}{dt^2}\frac{\partial L}{\partial \varphi'''} , \quad P_2 := \frac{\partial L}{\partial \varphi''} - \frac{d}{dt}\frac{\partial L}{\partial \varphi'''} \quad \text{and} \quad P_3 := \frac{\partial L}{\partial \varphi'''} . \text{ Eq.3.28}$$

We do not really need the third-order formalism to determine the Hamiltonians of the Lagrangians given in Eqs.3.15-3.18. This is so because the third-order Lagrangians, $L_4$ and $L_5$ presented in Eqs.3.17 and 3.18 differ from the second-order Lagrangians $L_{4SO}$ and $L_{5SO}$ by divergences, and as a result yield the same expressions for H, when $\varphi=\varphi(t)$.

One useful property of the Ostrogradsky Hamiltonian is that it satisfies the following identity

$$\frac{dH}{dt} = -\varphi' \frac{\delta L}{\delta \varphi} \quad . \qquad\qquad\qquad \text{Eq.3.29}$$

Thus, as is well-known, when the field equations are satisfied, H is a constant, and conversely, when H is constant, the field equations are satisfied. Thus the equation H= constant, is a first integral of our field equations. In particular we shall be interested in the solutions to H = 0, which I shall call the vacuum solutions to our empty space scalar-scalar theory. This equation will be expressed in terms of $\varphi$ and its derivatives. I have no interest in using the Legendre transformation (*see, e.g.,*[11]) to re-express H in terms of the canonical variables.

At this point I should remark that some people may regard it as inappropriate for me to refer to the solutions of H = 0, as the vacuum solutions, since that term is often reserved for the solution to H = constant, which is of lowest energy. In Appendix A I show that there exists a scalar-scalar theory for which the equation H = constant, admits solutions for all choices of that constant. And so for that theory there is no minimum energy solution. Nevertheless, for the remainder of this paper I shall refer to the solutions to H = 0 as vacuum solutions. We shall now begin our quest for such solutions.

It is a straightforward matter to employ Eqs.3.15, 3.16, 3.19, 3.20, 3.27 and 3.28 to show that

$$H_2 = 2 K(\varphi')^3 - 2 K_{,X} (\varphi')^5 \qquad\qquad \text{Eq.3.30}$$

$$H_3 = 4 G_{3,\varphi}(\varphi')^5 \qquad\qquad \text{Eq.3.31}$$

$$H_4 = 12 G_4 (\varphi')^2\varphi''' + 24 G_{4,X}(\varphi')^4\varphi''' + 12 G_{4,X}(\varphi')^3(\varphi'')^2 + 12 G_{4,\varphi}(\varphi')^3\varphi'' +$$
$$+ 6 G_{4,\varphi\varphi}(\varphi')^5 + 24 G_{4,X\varphi}(\varphi')^5\varphi'' - 24 G_{4,XX}(\varphi')^5(\varphi'')^2 \qquad \text{Eq.3.32}$$

$$H_5 = 6 G_{5,\varphi}(\varphi')^4\varphi''' + 6 G_{5,\varphi}(\varphi')^3(\varphi'')^2 + 12 G_{5,X}(\varphi')^4\varphi''\varphi''' + 8 G_{5,X}(\varphi')^3(\varphi'')^3 +$$
$$+ 6G_{5,\varphi\varphi}(\varphi')^5\varphi'' - 8 G_{5,XX}(\varphi')^5(\varphi'')^3 . \qquad \text{Eq.3.33}$$

For the present form of our metric and scalar field the Lagrangian $L_3$ and its corresponding Hamiltonian $H_3$ can be assimilated into $L_2$ and $H_2$. To see why let us choose K to be

$$K := -2X^{-1} \int XG_{3,\varphi}(\varphi,X)dX .$$

Using this value of K in Eq.3.30 we end up with the Hamiltonian $H_3$ given in Eq.3.31, which corroborates my claim that $L_3$ can be assimilated into $L_2$ for our present purposes.

I shall set $L_{2,4} := L_2 + L_{4SO}$ and $L_{2,5} := L_2 + L_{5SO}$, with the corresponding

Hamiltonians $H_{2,4} = H_2 + H_4$ and $H_{2,5} = H_2 + H_5$ . Recall that we are expressing our Hamiltonians in terms of $\varphi$ and its time derivatives, and not in terms of canonical variables. This justifies my expressions for $H_{2,4}$ and $H_{2,5}$. I shall now show that the coefficient functions appearing in $H_{2,4}$ can be chosen so that the equation $H_{2,4}=0$ admits solutions of the form $e^{\beta t}$ and $t^q$ ($\beta$ and q some real numbers), with the coefficient functions in $H_{2,4}$ for these two different classes of solutions being unchanged. Note this differs from the situation with $L_s$ above, where different choices of K were needed to get the solutions presented in Eqs.3.25 and 3.26. After showing how the equation $H_{2,4} = 0$, can be solved, I shall present solutions to $H_{2,5} = 0$.

We shall seek solutions to $H_{2,4} = 0$, when K and $G_4$ have the form

$$K := A_2\, \varphi^\mu\, |X|^\nu = A_2\, \varphi^\mu(\varphi')^{2\nu}\ \text{ and } \ G_4 := A_4\, \varphi^\eta\, |X|^\zeta = A_4\varphi^\eta(\varphi')^{2\zeta}\,, \qquad\qquad \text{Eq.3.34}$$

where $A_2$, $A_4$, $\mu$, $\nu$, $\eta$ and $\zeta$ are numbers to be determined. To begin we place our ansatz expressions for K and $G_4$ into the equation $H_{2,4} = 0$. Using Eqs.3.30. 3.32 and 3.34 we find

$$\tfrac{1}{2}H_{2,4} \equiv\ A_2(\nu+1)\varphi^\mu(\varphi')^{2\nu+3} + 6A_4(1-2\zeta)\varphi^\eta(\varphi')^{2\zeta+2}\varphi''' +$$

$$+\ 6A_4\,\zeta(1-2\zeta)\varphi^\eta(\varphi')^{2\zeta+1}(\varphi'')^2 + 6A_4\,\eta(1-2\zeta)\varphi^{\eta-1}(\varphi')^{2\zeta+3}\varphi'' +$$

$$+\ 3A_4\eta(\eta-1)\varphi^{\eta-2}(\varphi')^{2\zeta+5}\,. \qquad\qquad\qquad \text{Eq.3.35}$$

We desire a solution to $H_{2,4} = 0$, of the form $\varphi = \alpha e^{\beta t}$, where $\alpha$ and $\beta$ are real constants with $\alpha \sim \ell^1$ and $\beta \sim \ell^{-1}$. Upon putting this form for $\varphi$ and its derivatives into Eq.3.35, we discover that in order for φ=αe^{βt} to be a solution to $H_{2,4}=0$, we must have

$$\mu + 2\nu = \eta + 2\zeta \qquad\qquad\qquad Eq.3.36$$

and

$$(\nu+1)A_2\beta^{(2\nu+3)} + A_4\beta^{2\zeta+5}[3\eta + 3\eta^2 + 6 - 12\eta\zeta - 6\zeta - 12\zeta^2] = 0 \qquad Eq.3.37$$

with α being an arbitrary real number. It should be noted that Eqs.3.35 and 3.36 imply that the equation $H_{2,4} = 0$ is a homogeneous differential equation.

Next we seek a solution to $H_{2,4}=0$ of the form

$$\varphi = \gamma(k_1 t + k_2)^q, \qquad\qquad\qquad Eq.3.38$$

where $\gamma\sim\ell^1$, $k_1\sim\ell^{-1}$, $k_2\sim\ell^0$ and $q\sim\ell^0$ are constants. (If we were working with unitless

quantities we could pull $k_1$ out of the expression $k_1 t+k_2$ in Eq.3.38, and absorb it into

γ and $k_2$. But we would need to replace $k_1$ by ε = ±1. In addition, since various

constants in Eq.3.38 have units, pulling $k_1$ out would cause γ to have very strange, and

possibly irrational units. That is why I left $k_1$ where it is.)  To determine this solution

we insert the expression for φ and its derivatives into Eq.3.35 and discover, after some

effort, that in order for Eq.3.38 to satisfy the equation $H_{2,4} = 0$, we must have

$$q\mu + 2q\nu - 2\nu = q\eta + 2q\zeta - 2\zeta - 2 . \qquad\qquad Eq.3.39$$

Eqs.3.36 and 3.39 combine to give us

$$\nu = \zeta + 1, \qquad\qquad\qquad Eq.3.40$$

for our present choices of K and $G_4$. This in turn implies that the β's in Eq.3.37 cancel

leaving us with

$$(\zeta+2)A_2 + A_4[3\eta + 3\eta^2 + 6 - 12\eta\zeta - 6\zeta - 12\zeta^2] = 0 \ . \qquad\qquad Eq.3.41$$

In addition Eqs.3.36 and 3.40 show that we must also have

$$\mu = \eta - 2 \ . \qquad\qquad Eq.3.42$$

When these values for $\mu$ and $\nu$ are placed into Eq.3.35, with $\varphi$ given by Eq.3.38, the $\gamma$ and $k_1$ terms can be eliminated, as well as a lot of common factors of q, leaving us with

$$(\zeta+2)A_2 q^2 + A_4[24\zeta q - 18\zeta - 6\zeta q^2 - 12\zeta\eta q^2 + 12\zeta\eta q - 6\eta q - 12\zeta^2 q^2 + 24\zeta^2 q +$$
$$- 12\zeta^2 + 6q^2 - 18q + 12 + 3\eta^2 q^2 + 3\eta q^2\ ] = 0 \ . \qquad\qquad Eq.3.43$$

If we multiply Eq.3.41 by $q^2$, and subtract that from Eq.3.43, we discover that all of the $q^2$ terms cancel, leaving us with

$$A_4[24\zeta q - 18\zeta + 12\zeta\eta q - 6\eta q + 24\zeta^2 q - 12\zeta^2 - 18q + 12] = 0. \qquad Eq.3.44$$

Since we are assuming that $A_4$ is non-zero, Eq.3.44 implies that

$$q = \frac{2\zeta^2 + 3\zeta - 2}{4\zeta^2 + 4\zeta + 2\zeta\eta - \eta - 3} = \frac{(2\zeta - 1)(\zeta + 2)}{(2\zeta - 1)(2\zeta + \eta + 3)} \ . \qquad\qquad Eq.3.45$$

So if $\zeta \neq \tfrac{1}{2}$ , then Eq.3.45 tells us that

$$q = \frac{\zeta + 2}{2\zeta + \eta + 3} \ , \qquad\qquad Eq.3.46$$

while if $\zeta = \tfrac{1}{2}$ then Eq.3.44 holds for any choice of q and $\eta$. If we place $\zeta = \tfrac{1}{2}$ into Eqs.3.41 and 3.43 we obtain the same relationship between $A_2$ and $A_4$; *viz.,*

$$A_2 = [6\eta(\eta - 1)A_4]/5 \ . \qquad\qquad Eq.3.47$$

In summary if we choose $\zeta = \frac{1}{2}$, then we shall obtain solutions to $H_{2,4} = 0$ of the form $\varphi = \alpha e^{\beta t}$ and $\varphi = \gamma(k_1 t + k_2)^q$, where $\alpha$, $\beta$, $\gamma$, $k_1$, $k_2$ and q are arbitrary provided we choose

$$K = A_2 \, \varphi^{\eta-2} \, |X|^{3/2} \text{ and } G_4 = A_4 \, \varphi^{\eta} \, |X|^{1/2} \qquad \text{Eq.3.48}$$

with $A_2$ and $A_4$ being related by Eq.3.47. This strikes one as remarkable until you examine the form of $H_{2,4}$ for this choice of $\zeta$, K and $G_4$, which is given by Eq.3.35. Upon doing so we find that $H_{2,4} \equiv 0$, which clearly explains why that equation admits so many solutions, it's 0 = 0! However, later we shall find a non-trivial Hamiltonian that does admit all the above solutions, and no others, as vacuum solutions.

Thus we can assume that $\zeta \neq \frac{1}{2}$. Consequently Eq.3.41 provides us with the relationship between $A_2$ and $A_4$. If $\zeta = -2$ then we see that $A_2$ is arbitrary, while due to Eq.3.46 q must vanish. This situation is uninteresting, so we shall now assume that $\zeta \neq -2$, and hence

$$A_2 = 3A_4 \frac{[\, 4\eta\zeta + 4\zeta^2 + 2\zeta - \eta^2 - \eta - 2 \,]}{(\zeta + 2)} . \qquad \text{Eq.3.49}$$

In summary, if we choose values of $\eta$, $\zeta$ and $A_4$ (with $\zeta \neq -\frac{1}{2}$ or $-2$, and $2\zeta + \eta + 3 \neq 0$), and use them to compute $\nu$, $\mu$, q and $A_2$ with Eqs.3.40, 3.42, 3.46 and 3.49, then $\varphi = \alpha e^{\beta t}$ and $\varphi = \gamma(k_1 t + k_2)^q$ will be solutions to $H_{2,4} = 0$ for all values of $\alpha$, $\beta$, $\gamma$, $k_1$ and $k_2$ for which $\varphi' > 0$. The selected values of $\eta$, $\zeta$ and $A_4$ can be used in Eq.3.34 to provide us with expressions for K and $G_4$.

For Horndeski theory it has been shown (*see,* Baker, *et al.,*[12], Creminelli & Vernizzi [13], Sakstein & Jain [14] and Ezquiaga & Zumalaćarregui [15]) that in order to guarantee that the speed of a gravitational wave, $c_g$, is equal to the speed of light, c, we need $G_{4,X} \approx 0$ and $G_5$ approximately constant. If you look at the derivation of this result in the references cited it involves a perturbation analysis of the metric tensor. Whether a similar result applies to the class of scalar-scalar theories I have presented here is not clear. But in any case let us look at our $H_{2,4} = 0$ solutions when $G_{4,X} = 0$, which requires $\zeta = 0$. In this case we can use Eqs.3.40, 3.42, 3.46 and 3.49 to conclude that $\mu = \eta - 2$, $\nu = 1$,

$$A_2 = -3A_4[\eta^2 + \eta + 2]/2 \quad \text{and} \quad q = 2(\eta + 3)^{-1} \ . \qquad \text{Eq.3.50}$$

Hence our Hamiltonian, which appears on the left-hand side of Eq.3.35, is

$$H_{2,4} = -12A_4(\eta+1)\varphi^{\eta-2}(\varphi')^5 + 12A_4\varphi^{\eta}(\varphi')^2\varphi''' + 12\eta A_4\varphi^{\eta-1}(\varphi')^3\varphi'' \ . \qquad \text{Eq.3.51}$$

I would like a solution for $\varphi$ of the form $\varphi = \gamma(k_1t + k_2)^q$ which is such that $\varphi'(0) = 0$, and $\varphi'(t)$ has a vertical tangent vector at t = 0 (more precisely we want $\varphi'(t) \to 0$, and $\varphi''(t) \to \infty$, as $t \to 0^+$). In view of the form of our metric given in Eq.3.3 this would correspond to a universe that expands infinitely fast at t=0. None of our exponential solutions can generate such a universe since they never have vertical tangent vectors. Nevertheless they turn out to be very useful, as we shall see in the next section.

To obtain a $\varphi$ solution of the type I described in the above paragraph we need

$k_2 = 0$, and q to lie in the range $1 < q < 2$. In view of Eq.3.50 we see that this restriction on q implies that $-2 < \eta < -1$. The question now is: what should we choose as reasonable value of q between 1 and 2? Well, when people look for inflationary models of the universe they seek models in which the curvature tensor rapidly vanishes as the universe expands. Our present model has the $t = $ constant slices being flat, and it is easily seen from Eq.3.11 that the scalar curvature, R, for this model is given by

$$R = 6(q-1)(2q-3)t^{-2} , \qquad\qquad Eq.3.52$$

which, serendipitously, vanishes for $q = 3/2$, which is in the middle of our admissible range for q. For this value of q, $\eta = -5/3$, and hence the coefficient functions K and $G_4$ appearing in $L_{2,4}$ are

$$K = (-14A_4/3)\varphi^{-11/3}|X|, G_4 = A_4\varphi^{-5/3} .$$

One should also note that this solution is not flat for all time, since the Ricci tensor components are non-zero and go as $t^{-2}$. In the next section I shall discuss how the explosive solution to $H_{2,4} = 0$, can be combined with the exponential solutions to yield cosmological models which are strikingly similar to the usual inflationary model.

I shall now make a few remarks about the vacuum solutions to the scalar-scalar theory generated by the Lagrangian $L_{2,5} = L_2 + L_{5SO}$.

If we take $K = A_2\varphi^\mu|X|^\nu$ and $G_5 = A_5\varphi^\chi|X|^\xi$, then it is a straightforward matter to use Eqs.3.30 and 3.33 to show that

$$\tfrac{1}{2}H_{2,5} = (1+\nu)A_2\varphi^\mu(\varphi')^{2\nu+3} + 3\chi A_5\varphi^{\chi-1}(\varphi')^{2\xi+3}(\varphi'')^2 - 4\xi^2 A_5\varphi^\chi(\varphi')^{2\xi+1}(\varphi'')^3 +$$

$$+ 3\chi(\chi-1)A_5\varphi^{\chi-2}(\varphi')^{2\xi+5}\varphi'' + 3\chi A_5\varphi^{\chi-1}(\varphi')^{2\xi+4}\varphi''' - 6\xi A_5\varphi^\chi(\varphi')^{2\xi+2}\varphi''\varphi''' . \quad \text{Eq.3.53}$$

For $\varphi = \alpha e^{\beta t}$ to be a solution to $H_{2,5} = 0$, we must have

$$\mu + 2\nu = \chi + 2\xi + 1 , \qquad\qquad\qquad\qquad \text{Eq.3.54}$$
and
$$A_2(\nu+1)\beta^{2\nu} + A_5\beta^{2\xi+4}[3\chi + 3\chi^2 - 4\xi^2 - 6\xi] = 0 . \qquad \text{Eq.3.55}$$

As was the case for the equation $H_{2,4}=0$, the equation $H_{2,5}=0$, must be a homogeneous differential equation, due to Eq.3.54.

Next we seek a second solution to $H_{2,5} = 0$ of the form $\varphi = \gamma(k_1 t + k_2)^q$. In order for a solution of this form to exist along with solutions of the form $\varphi = \alpha e^{\beta t}$ we must have

$$q\mu + 2q\nu - 2\nu = q\chi + 2q\xi + q - 2\xi - 4 .$$

Due to Eq.3.54 this equation simplifies to

$$\nu = \xi + 2 , \qquad\qquad\qquad\qquad\qquad \text{Eq.3.56}$$

which can be combined with Eq.3.54 to give us

$$\mu = \chi - 3 . \qquad\qquad\qquad\qquad\qquad \text{Eq.3.57}$$

Eq.3.56 implies that the $\beta$ terms in Eq.3.55 cancel out leaving us with

$$(\nu + 1)A_2 + A_5[3\chi^2 + 3\chi - 4\xi^2 - 6\xi] = 0 . \qquad\qquad \text{Eq.3.58}$$

Eq.3.56 also permits us to cancel the $\gamma$ and $k_1$ terms that arise when we evaluate $H_{2,5}=0$ for $\varphi = \gamma(k_1 t + k_2)^q$, giving us

$$(\nu+1)A_2q^3 + A_5[3\chi q^3 - 12\chi q^2 + 9\chi q - 4\xi^2 q^3 + 12\xi^2 q^2 - 12\xi^2 q + 4\xi^2 + 3\chi^2 q^3 +$$

$$- 3\chi^2 q^2 - 6\xi q^3 + 24\xi q^2 - 30\xi q + 12\xi] = 0 . \qquad \text{Eq.3.59}$$

If we multiply Eq.3.58 by $q^3$ and subtract the resulting equation from Eq.3.59 we obtain

$$A_5[-12\chi q^2 + 9\chi q + 12\xi^2 q^2 - 12\xi^2 q + 4\xi^2 - 3\chi^2 q^2 + 24\xi q^2 - 30\xi q + 12\xi] = 0 , \quad \text{Eq.3.60}$$

which is a quadratic equation relating q, $\chi$ and $\xi$.

Now recall what we are trying to do here. We want to choose the exponents $\mu$, $\nu$, $\chi$ and $\xi$ along with the coefficients $A_2$ and $A_5$ appearing in our expressions for K and $G_5$; *viz.,*

$$K = A_2\varphi^\mu|X|^\nu \quad \text{and} \quad G_5 = A_5\varphi^\chi|X|^\xi \qquad \text{Eq.3.61}$$

so that the resulting Hamiltonian vacuum equation $H_{2,5} = 0$, admits simultaneous solutions of the form $\varphi = \alpha e^{\beta t}$ and $\varphi = \gamma(k_1 t + k_2)^q$. This can be achieved by choosing values of $\chi$ and $\xi$ , which can then be used in Eqs 3.56 and 3.57 to give us values for $\mu$ and $\nu$. The values of $\chi$ and $\xi$ are then used in Eqs.3.58 and 3.60 to obtain values for $A_2$ and q. In the solutions so obtained $\alpha$, $\beta$, $\gamma$, $k_1$ and $k_2$ are only constrained by the condition that $\varphi' > 0$.

To illustrate this approach to obtaining solutions to $H_{2,5} = 0$, let us choose $\xi = 0$, and leave the choice of $\chi$ temporarily unfixed. Under this assumption Eq.3.60 tells us that q must satisfy

$$4\chi q^2 - 3\chi q + \chi^2 q^2 = 0. \qquad\qquad\qquad \text{Eq.3.62}$$

If $\chi = 0$ in Eq.3.62, then we can employ Eqs.3.53, 3.56 and 3.58 to deduce that $H_{2,5} \equiv 0$, which is undesirable. Hence $\chi \neq 0$, and Eq.3.62 tells us that

$$q = 3(4 + \chi)^{-1}. \qquad\qquad\qquad \text{Eq.3.63}$$

To obtain solutions of the form $\varphi = \gamma(k_1 t + k_2)^q$ which begin explosively at $t = 0$, we require $k_2 = 0$, with $1 < q < 2$, and hence Eq.3.63 implies that $-2.5 < \chi < -1$. If we would also wish $R = 0$ for our explosive solution, then we would need to choose $q = 1.5$ and $\chi = -2$. From Eq.3.52 we know that for fixed values of $t$, the minimum value of $R$ is attained when $q = 1.25$, which would require us to choose $\chi = -{}^8/_5$. The values for $A_2$ and $A_5$ for each of these solutions can be found from Eq.3.58. When $q = 1.5$ we have $A_2 = -2A_5$, and when $q = 1.25$ we have $A_2 = -24A_5/25$.

When we dealt with the $H_{2,4} = 0$ equation above, we saw that for solutions of the form $\varphi = \gamma(k_1 t + k_2)^q$ to exist, $q$ must be given by Eq.3.46 in terms of $\eta$ and $\zeta$, with only one $q$ value corresponding to each choice of $\eta$ and $\zeta$. I shall now demonstrate that when dealing with the equation $H_{2,5} = 0$, we can choose $\chi$ and $\xi$ in $G_5 = A_5 \varphi^\chi |X|^\xi$, so that solutions of the form $\varphi = \gamma(k_1 t + k_2)^q$ can exist for two different choices of $q$.

To that end let us choose $\chi = 0$. Then Eq.3.60 tells us that

$$\xi^2(12q^2 - 12q + 4) + \xi(24q^2 - 30q + 12) = 0. \qquad\qquad \text{Eq.3.65}$$

$\xi \neq 0$ in this equation, since if it did $H_{2,5}$ would reduce to $H_2$, due to Eq.3.53, which is

not what we wish to investigate. Thus Eq.3.65 reduces to

$$\xi(6q^2 - 6q + 2) + 12q^2 - 15q + 6 = 0 ,$$   Eq.3.66

and hence

$$\xi = \frac{15q - 12q^2 - 6}{6q^2 - 6q + 2} .$$   Eq.3.67

Say we choose $q = {}^3/_2$. Then Eq.3.67 tells us that $\xi = {}^{-21}/_{13}$. If we put this value of $\xi$ into Eq.3.65 we find that $q = {}^3/_2$ or ${}^4/_5$. Using Eqs.3.56-3.58 we also find that $(\mu, \nu) = (-3, {}^5/_{13})$ and $A_2 = 7A_5/13$.

If we look for a solution with $q = {}^5/_4$, then following the above approach we obtain $\xi = {}^{-48}/_{31}$. This in turn gives rise to two values of q, *viz.*, $q = {}^5/_4$, and ${}^6/_7$, with $(\mu, \nu) = (-3, {}^{14}/_{31})$, and $A_2 = 608A_5/155$.

I shall now summarized what we have just done with the Hamiltonian $H_{2,5}$. The equation $H_{2,5} = 0$, with $K := A_2\varphi^\mu|X|^\nu$ and $G_5 := A_5\varphi^\chi|X|^\xi$, will admit simultaneous solutions of the form $\varphi = \alpha e^{\beta t}$ and $\varphi = \gamma(k_1t + k_2)^q$, with $\alpha$, $\beta$, $\gamma$, $k_1$ and $k_2$ arbitrary, except that $\varphi'>0$, when:

(i)    $(\mu, \nu) = (-5, 2)$,   $(\chi, \xi) = (-2, 0)$, $A_2 = -2A_5$, $q = {}^3/_2$ ;

(ii)    $(\mu, \nu) = ({}^{-23}/_5, 2)$,  $(\chi, \xi) = ({}^{-8}/_5, 0)$, $A_2 = -24A_5/25$, $q = {}^5/_4$;

(iii)    $(\mu, \nu) = (-3, {}^5/_{13})$, $(\chi, \xi) = (0, {}^{-21}/_{13})$, $A_2 = 7A_5/13$, $q = {}^3/_2$ or ${}^4/_5$; and

(iv)    $(\mu, \nu) = (-3, {}^{14}/_{31})$,  $(\chi, \xi) = (0, {}^{-48}/_{31})$, $A_2 = 608A_5/155$, $q = {}^5/_4$ or ${}^6/_7$.

For our last example we shall consider conformally invariant scalar-tensor field theories. If L is the Lagrangian of a scalar-tensor field theory, then that theory will be said to be conformally invariant, if the field tensor densities

$$g_{bc}\frac{\delta L}{\delta g_{ac}} \quad \text{and} \quad \frac{\delta L}{\delta \varphi}$$

are invariant under the conformal transformation which replaces $g_{ij}$ by $e^{2\sigma}g_{ij}$ throughout the field tensors, with $\sigma$ being an arbitrary scalar field. The scalar-tensor theory will also be said to be flat space compatible if L is well-defined and differentiable when evaluated for either a flat metric tensor, or constant scalar field. In [16] (and also in [17]) I show that in a 4-dimensional space the Lagrangian of the most general conformally invariant scalar-tensor field theory which is flat-space compatible is given by

$$L_C = L_{2C} + L_{3C} + L_{4C} + L_{UC} \qquad\qquad\qquad \text{Eq.3.68}$$

where

$$L_{2C} := g^{1/2}K(\varphi)X^2 , \qquad\qquad\qquad\qquad \text{Eq.3.69a}$$

$$L_{3C} := P(\varphi)\varepsilon^{abcd}C^{ef}{}_{ab}C_{efcd} \qquad\qquad\qquad \text{Eq.3.69b}$$

$$L_{4C} := g^{1/2}B(\varphi)C^{abcd}C_{abcd} \qquad\qquad\qquad \text{Eq.3.69c}$$

and

$$L_{UC} := g^{1/2}U(\varphi)[-12R^{ab}\varphi_a\varphi_b + 2RX - 3(\Box\varphi)^2 - 6\varphi^{ab}\varphi_{ab} - 12\varphi^a(\Box\varphi)_{,a}] \quad \text{Eq.3.69d}$$

where

$$C^{abcd} := R^{abcd} + \tfrac{1}{2}(g^{ad}R^{bc} + g^{bc}R^{ad} - g^{ac}R^{bd} - g^{bd}R^{ac}) + \tfrac{1}{6}R(g^{ac}g^{bd} - g^{ad}g^{bc})$$

are the components of the Weyl tensor, K, P, B and U are differentiable functions of $\varphi$, and $\varepsilon^{abcd}$ is the Levi-Civita tensor density.

We shall now seek solutions to the scalar-scalar theory based on the Lagrangian

$L_C$ with the metric tensor given by Eq.3.3 and corresponding line element

$$ds^2 = -dt^2 + (\varphi')^2[du^2 + dv^2 + dw^2] = (\varphi')^2[-(\varphi'^{-1}dt)^2 + du^2 + dv^2 + dw^2] \ . \quad \text{Eq.3.70}$$

We see if we define a new coordinate $\bar{t}$ by $d\bar{t} := \varphi'^{-1}dt$, then it is obvious that our metric is conformally flat. Consequently when we evaluate $L_{3C}$ and $L_{4C}$ for our ansatz metric they vanish, which certainly makes life a lot easier. So we can employ Eqs.3.9-3.14 to deduce that

$$L_C = K(\varphi')^7 \ - \ 12U[2(\varphi')^4\varphi''' + 7(\varphi')^3(\varphi'')^2],$$

which can be rewritten as

$$L_C = K(\varphi')^7 + 24U_\varphi(\varphi')^5\varphi'' + 12U(\varphi')^3(\varphi'')^2 - \underline{d}(24U(\varphi')^4\varphi'') \ . \qquad \text{Eq.3.71}$$
$$\phantom{L_C = K(\varphi')^7 + 24U_\varphi(\varphi')^5\varphi'' + 12U(\varphi')^3(\varphi'')^2 - }\overline{dt}$$

Using Eq.3.27 and 3.71 we find that the Ostrogradsky Hamiltonian associated with $L_C$ is given by

$$H_C = -24(\varphi')^3 \left[ \ \underline{d}(U\varphi'\varphi'') \ - \ [\tfrac{1}{4}K \ - \ U_{\varphi\varphi}](\varphi')^4 \right] \ . \qquad \text{Eq.3.72}$$
$$\phantom{H_C = -24(\varphi')^3 \left[ \ \right.}\overline{dt}$$

The vacuum solutions for the scalar-scalar theory based on the Lagrangian $L_C$ are the solutions to $H_C = 0$. Since we require that $\varphi' > 0$, we can use Eq.3.72 to deduce that the vacuum solutions for $\varphi$ must satisfy

$$\underline{d} \ (U\varphi'\varphi'') = [\tfrac{1}{4}K - U_{\varphi\varphi}](\varphi')^4 \ . \qquad\qquad \text{Eq.3.73}$$
$$\overline{dt}$$

Let us proceed to construct solutions to $H_C = 0$, which is equivalent to solving Eq.3.73. We shall seek solutions similar to those that we found for $H_{2,4} = 0$, and $H_{2,5}$

= 0. To that end I shall assume that $K := A\varphi^a$ and $U := B\varphi^b$, where A, B, a and b are real numbers to be determined so that Eq.3.73 admits simultaneous solutions of the form $\varphi = \alpha e^{\beta t}$ and $\varphi = \gamma(k_1 t + k_2)^q$. Upon setting $\varphi = \alpha e^{\beta t}$ in Eq.3.73 it is easily shown that this expression for $\varphi$ is a solution to Eq.3.73 if and only if

$$a = b - 2 \text{ , and } A = 4(b^2 + 2)B \text{ ,} \hspace{3cm} \text{Eq.3.74}$$

and hence

$$K = 4(b^2 + 2)B\varphi^{b-2} \text{ , and } U = B\varphi^b \text{ .} \hspace{2cm} \text{Eq.3.75}$$

Once again Eqs.3.74 and Eq.3.75 force the equation $H_C = 0$, as formulated in Eq.3.73, to be a homogeneous differential equation.

Next we shall seek a solution to Eq.3.73 of the form $\varphi = \gamma(k_1 t + k_2)^q$, under the assumption that Eq.3.74 holds. It is easily seen that these conditions imply that $\varphi$ will be a solution for arbitrary choices of $\gamma$, $k_1$ and $k_2$ provided

$$q = 3/(5 + b) \text{ .} \hspace{4cm} \text{Eq.3.76}$$

In order to obtain solutions with $1 < q < 2$, we need $^{-7}/_2 < b < -2$. So, for example, if we choose $q = {}^3/_2$, in our expression for $\varphi = \gamma(k_1 t + k_2)^q$, so that the scalar curvature, R, vanishes for this solution, then Eqs.3.75 and 3.76 tell us that

$$a = -5, b = -3 \text{ and } A = 44B \text{ .}$$

Consequently to guarantee that $\varphi = \alpha e^{\beta t}$, and $\varphi = \gamma(k_1 t + k_2)^{3/2}$ are solutions to $H_C = 0$, with $\alpha$, $\beta$, $\gamma$, $k_1$ and $k_2$ arbitrary (except for $\varphi' > 0$), we require

$K = 44B\varphi^{-5}$ and $U = B\varphi^{-3}$ . $\qquad\qquad$ Eq.3.77

In the next section I shall explain how the various solutions to the scalar-scalar theories that we have constructed here can be used to generate cosmological models. But before doing that another remark is in order.

It is surprising that the hidden scalar Lagrangians in scalar-tensor Lagrangians as different as $L_{2,4}$, $L_{2,5}$ and $L_C$ can give rise to simultaneous solutions of the form $\varphi = \alpha e^{\beta t}$ and $\varphi = \gamma(k_1 t + k_2)^{3/2}$, when the coefficient functions in these Lagrangians are chosen suitably. The reason this is so, stems from the fact that the Hamiltonians associated with these three Lagrangians are closely related when the coefficient functions are chosen to give rise to these simultaneous solutions for $\varphi$. In fact it is not difficult to demonstrate, using Eqs.3.51 (with $\eta = -{}^5/_3$), 3.53 (with $\mu = -5$, $\nu = 2$, $\chi = -2$, $\xi = 0$, $A_2 = -2A_5$), 3.72 and 3.77 that

$H_C = 2BH_{2,5}/A_5$ and $mH_{2,4} + H_{2,5} + nH = 0,$ $\qquad$ Eq.3.78

where

$H_C = -24B(\varphi')^3[\varphi^{-5}(\varphi')^4 - 3\varphi^{-4}(\varphi')^2\varphi'' + \varphi^{-3}(\varphi'')^2 + \varphi^{-3}\varphi'\varphi''']$ $\qquad$ Eq.3.79

$H_{2,4} = A_4[8\varphi^{-11/3}(\varphi')^5 - 20\varphi^{-8/3}(\varphi')^3\varphi'' + 12\varphi^{-5/3}(\varphi')^2\varphi''']$ , $\qquad$ Eq.3.80

$H_{2,5} = A_5[-12\varphi^{-5}(\varphi')^7 + 36\varphi^{-4}(\varphi')^5\varphi'' - 12\varphi^{-3}(\varphi')^3(\varphi'')^2 - 12\varphi^{-3}(\varphi')^4\varphi''']$ , Eq.3.81

$H := -\kappa[(\varphi')^{-1}\varphi'' - 2\varphi(\varphi')^{-3}(\varphi'')^2 + \varphi(\varphi')^{-2}\varphi'''] = -\kappa\dfrac{d}{dt}[\varphi(\varphi')^{-2}\varphi'']$ , $\qquad$ Eq.3.82

$m := 1.5\varphi^{-4/3}(\varphi')^2A_5/A_4$ $\qquad\qquad$ Eq.3.83

$n := 6\varphi^{-4}(\varphi')^6A_5/\kappa$ , $\qquad\qquad$ Eq.3.84

and $\kappa$ is a dimensioned constant. The new quantity $H$ defined in Eq.3.82 is the Hamiltonian of the Lagrangian

$$L := \tfrac{1}{2}\kappa\varphi(\varphi')^{-3}(\varphi'')^2 .$$

<div align="right">Eq.3.85</div>

$L$ and $H$ are very interesting quantities. First of all the vacuum equation $H = 0$, can be completely solved, and the solutions are $\varphi = \alpha e^{\beta t}$, and $\varphi = \gamma(k_1 t + k_2)^q$, where $\alpha$, $\beta$, $\gamma$, $k_1$, $k_2$ and $q$ are arbitrary up to the requirement that $\varphi' > 0$ (so $q \neq 0$). Because of this fact, and Eq.3.78, it should now be clear why the three seemingly different theories determined by $L_{2,4}$, $L_{2,5}$ and $L_C$ converge when we require that each one has $\varphi$ solutions of the form $\varphi = \alpha e^{\beta t}$ and $\varphi = \gamma(k_1 t + k_2)^{3/2}$. So when we seek solutions of this form, we are not really dealing with three separate scalar-scalar theories, but only one. The choice is yours.

The next interesting property of $L$ and $H$ is that $L$ is the hidden scalar Lagrangian of a "true" scalar-scalar field theory. To see this we define a new scalar field on our spacetime M by $f := f(d\varphi)$. When $\varphi = \varphi(t)$, and f has the FLRW form given by Eq.2.7, then $f = X$. But in general $\varphi$ and f need not have these forms. Using $f$ we define a scalar-scalar Lagrangian $\mathcal{L}_f$ by

$$\mathcal{L}_f := -\tfrac{1}{8}\kappa g^{\frac{1}{2}}\, \varphi\, f^{-4}\, g^{ab} f_{,a} f_{,b} .$$

<div align="right">Eq.3.86</div>

It is easily seen that when $\mathcal{L}_f$ is evaluated for the FLRW scalar-scalar theory which we have been studying, it gives rise to the Lagrangian $L$ given in Eq.3.85.

A scalar-tensor Lagrangian $\mathcal{L}_{ST}$ which gives rise to $L$ as its hidden Lagrangian when working with the FLRW scalar-scalar theory is

$$\mathcal{L}_{ST} := -\tfrac{1}{8}\kappa g^{\frac{1}{2}}\varphi X^{-4}g^{ab}X_{,a}X_{,b}. \qquad\qquad\qquad \text{Eq.3.87}$$

However, $\mathcal{L}_f$ is more general than $\mathcal{L}_{ST}$ when working in the context of scalar-scalar theories since $f \neq X$ in general. $\mathcal{L}_f$ also permits us to introduce terms into f of the form $\xi^i y_i + \psi$, where $\xi^i$ and $\psi$ are functions of $\chi^j$. Such "gauge transforming" terms do not directly effect $g^{ij}$, but they do effect $\mathcal{L}_f$, and hence the field equations.

In concluding this section I would like to mention that in Appendix A, I determine all of the solutions to $H_{2,4} = 0$, and show that there exists five independent families of solutions to this equation. I also construct all of the solutions to $H = $ constant $:= 2\kappa\kappa_1 \neq 0$, and demonstrate that there exists three independent families of solutions. For two of these families $\varphi$ and $H$ have the same sign, while for the third family the relationship between the sign of $\varphi$ and $H$ depends upon whether the domain of $\varphi$ is unbounded (signs the same) or bounded (signs opposite). If we interpret $\varphi$ as dark energy, then it is nice to know that there is a relationship between the sign of $\varphi$ and the sign of $H$, which we think of as the total energy of the system.

## Section 4: Delineating Future Tasks and Constructing Universes

As I mentioned earlier, in [12], [13], [14] and [15] it was pointed out that, in the context of Horndeski Theory, to guarantee that $c = c_g$ we need $G_{4,X} \approx 0$ and $G_5$ to be approximately a constant. This fact was arrived at using perturbation analysis of the metric tensor in those scalar-tensor theories. We now need a similar analysis to be performed on those scalar-scalar theories based upon the Lagrangians $L_{2,4}$, $L_{2,5}$ and $L_C$, with corresponding Hamiltonians given in Eqs.3.35, 3.53 and 3.72. Completing this task in general might be difficult, in which case it would be nice to know what the relationship is between $c$ and $c_g$ in spaces which are given by some of the particular solutions to $H_{2,4} = 0$, $H_{2,5} = 0$ and $H_C = 0$ constructed in **Section 3**. In attempting to complete this task the first thing that needs to be determined is how to go about doing perturbation analysis in scalar-scalar theories. My suggestion is that since we fixed the choice of the Cofinsler function f in all the spaces constructed in the previous section, we should do the same thing when doing a perturbation analysis, and only vary $\varphi$.

The solutions to scalar-scalar field theories that we obtained in **Section 3** described empty space. How does one go about describing a more realistic situation involving matter? The easiest way to do this would be to adjoin to the Lagrangians $L_{2,4}$, $L_{2,5}$ and $L_C$ a matter Lagrangian $\mathcal{L}_M = \mathcal{L}_M(g_{ij}(d\varphi); g_{ij}(d\varphi)_{,h}; \Psi_A; \Psi_{A,h})$, where the $\Psi_A$'s represent the matter fields. Now just compute the Euler-Lagrange equations as we did in **Section 3**, and search for solutions.

In **Section 3** we confined our attention to FLRW spaces which were such that the

t = constant slices were flat.  It would be of interest to redo all of the calculations of **Section 3** for the cases in which the t = constant slices were either of constant positive or constant negative curvature.

When doing vector-tensor theory, the vector field is usually taken to be a covariant vector, $A = A_i dx^i$.  *E.g.,* this is the case in the Einstein-Maxwell field theory or my generalization of it [18].  Thus given a Lorentzian Cofinsler Space $CF^4 =$ (M,N,f), one can use A to construct a Lorentzian metric $g_A$ on M, provided that the range of A lies in N.  When A has this property, and x is any chart of M with corresponding standard chart $(\chi, y)$ in T*M, then the x components of this metric tensor are given by

$$g_A{}^{ij} := g_A(dx^i, dx^j) = \left[ \frac{\frac{1}{2}\partial^2 f(A)}{\partial y_i \partial y_j} \right].$$

One can evaluate Lagrangians of vector-tensor field theories using this $g^{ij}$ and $A_i$ to obtain vector-scalar Lagrangians, which can be used to generate field equations in a manner analogous to what we did in the previous section.

In the standard description of inflationary models of the Universe (*see,* Guth [19]), the Universe expands a little after t = 0 and stops, to achieve thermodynamic equilibrium, and then expands rapidly in an exponential manner  for a bit, after which there is  a moderate, albeit somewhat accelerated, expansion for eternity. I shall demonstrate that we can  obtain something like that using the solutions generated in the previous section. I shall also show that we can get the solutions to stop expanding, and then deaccelerate to $\varphi'(t) = 0$, at some time in the future.

I shall begin with some remarks upon the $\varphi' = 0$ solution. The equations $H_{2,4} = 0$, $H_{2,5} = 0$, and $H_C = 0$ all admit the spurious solution $\varphi' = 0$. We dispensed with this solution because in order for the metric given in Eq.3.3 to be non-degenerate we need $\varphi' \neq 0$. However, what if we took the $\varphi' = 0$ solution seriously. It simply means that in the slices t = constant, the physical distance between any two distinct points is zero, even though their Euclidean distance in these slices is non-zero. It is the physical distance between the constituents of the t = constant slices that enters into the calculations of all the non-gravitational forces between particles in these slices. So when $\varphi' = 0$ all of the particles in the t = constant slices are effectively contiguous, and thus you would expect that the strong force would vanish due to the asymptotic flatness of that force in more normal circumstances. As for the electric force, classically that force falls off as one over distance squared between charge particles, and hence would be infinite, unless the coupling constant vanished when $\varphi' = 0$. This leads me to suspect that when $\varphi' = 0$, all the non-gravitational forces between particles in a t=constant slice must vanish. Hence it is hard to see how thermodynamic equalibrium can be obtained in the slices where $\varphi' = 0$, if there are no forces acting, unless all particles have equal energy as an initial condition. However, we have reached this impasse by trying to discuss the behavior of large amounts of matter in solutions to equations which were source-free to begin with. The best we could do would be to discuss the behavior of test particles in the geometries we have obtained. Consequently for our (source-free) scalar-scalar theories based upon the Lagrangians investigated in **Section 3**, we really cannot say anything about whether thermodynamic equilibrium exists before the universe even begins to expand, because we do not have any matter in the Universe. In what follows I shall just use the $\varphi' = 0$ solution as a tool to bridge together other non-trivial solutions. In passing it should also be noted that $\varphi'=0$ is not a solution to the equations $H_{2,4}=$ constant, $H_{2,5}=$ constant, and $H_C=$ constant, when that constant is non-zero.

In **Section 3** we sought solutions to our field equations of the form $\varphi = \alpha e^{\beta t}$ and $\varphi = \gamma(k_1 t + k_2)^q$, for which $\varphi' > 0$. It is natural to at first restrict one's attention to the case where $\alpha>0$, $\beta>0$, $\gamma>0$ and $k_1>0$ (when q>0), since these give rise to expanding universes. However, the cases where $\alpha<0$, $\beta<0$, $\gamma<0$ and $k_1<0$ are also interesting, since they represent collapsing universe solutions, with $\varphi'>0$.

To illustrate how expanding and collapsing universes can be combined, let us consider some of the solutions to $H_{2,4} = 0$ obtained in the preceding section. If we want the space with $\varphi = \gamma(k_1 t + k_2)^q$ to have scalar curvature R = 0, we need to have $q = {}^3/_2$. Due to Eq.3.50, and the remarks following Eq.3.52, we see that if we choose K and $G_4$

in $L_{2,4}$ by

$$K = -14A_4\varphi^{-11/3}|X|/3 \quad \text{and} \quad G_4 = A_4\varphi^{-5/3} \quad , \qquad\qquad \text{Eq.4.1}$$

where $A_4$ is an arbitrary non-zero real number, then $\varphi = \alpha e^{\beta t}$ and $\varphi = \gamma(k_1 t + k_2)^{3/2}$ are solutions to $H_{2,4} = 0$, ($\forall$ choice of $\alpha$, $\beta$, $\gamma$, $k_1$ and $k_2$ which are such that $\varphi'>0$), and hence solutions to the Euler-Lagrange equation associated with $L_{2,4}$. So let us assume that our universe begins at $t = 0$ as $\varphi = \gamma(k_1 t + k_2)^{3/2}$ with $\varphi'(0) = 0$.  Thus we need $k_2 = 0$, and so $\varphi' = (3\gamma k_1/2)(k_1 t)^{\frac{1}{2}}$, with  $\varphi'' = (3\gamma k_1^2/4)(k_1 t)^{-\frac{1}{2}}$.  Unsurprisingly this solution begins explosively at $t = 0$, with a vertical tangent vector for the graph of $\varphi'$ versus t. As time increases the graph of $\varphi'$ passes through the graphs of $\varphi'$ for a myriad of solutions for which $\varphi = \alpha e^{\beta t}$, with $\varphi'=\alpha\beta e^{\beta t}$, where $\alpha>0$, and $\beta>0$.  From Eq.3.11 we see that for these exponential solutions $R = 12\beta^2$, and hence is never 0.  However, when $\beta>0$ is very close to 0, the solution $\varphi = \alpha e^{\beta t}$ , gives rise to  $\varphi' = \alpha\beta e^{\beta t}$, which stays "fairly flat," for a range of $t >0$,  although it continues an upward climb at an accelerating rate.  My contention is that when our original universe, with $\varphi' = (3\gamma k_1/2)(k_1 t)^{\frac{1}{2}}$, crosses a universe with $\varphi'$

$= \alpha\beta e^{\beta t}$ at some time $t_1$, it can "jump" from the original $\varphi$ state, to the exponential $\varphi$

state,  when $(3\gamma k_1/2)(k_1 t_1)^{\frac{1}{2}} = \alpha\beta \exp(\beta t_1)$, which, due to Eq.3.3, guarantees that the metric of the universe is continuous across the jump.  However, the first and second derivatives of the metric, in particular the scalar curvature R, are not continuous across the jump at time $t = t_1$, since R jumps from $R = 0$, to $R = 12\beta^2$.

Now the natural question to ask is how does the initial exploding state pick which exponential state it wants to jump into when the $\varphi$''s are equal? Well, one way to avoid this problem of making choices is to assume that everything that can happen does; *i.e.*, we are dealing with a multiverse. I shall make this thought a bit more precise. Suppose that $\forall$ choice of non-negative $\alpha$, $\beta$, $\gamma$, $k_1$ and $k_2 \in \mathbb{R}$, we define $\varphi_{\alpha,\beta} := \alpha e^{\beta t}$, and $\varphi_{\gamma,k1,k2} := \gamma(k_1 t + k_2)^{3/2}$, where $\varphi'_{\alpha,\beta} > 0$, and $\varphi'_{\gamma,k1,k2} > 0$. Let $\boldsymbol{Q}(^3/_2)$ denote the subset of $(\mathbb{R}^+)^4 \, x\mathbb{R}$ consisting of all $(\alpha, \beta, \gamma, k_1, k_2)$ which is such that the equation $\varphi'_{\alpha,\beta}(t) = \varphi'_{\gamma,k1,k2}(t)$ admits two solutions, and let $t_1$ be the smallest solution, and $t_2$ the second solution. Due to the nature of the exponential function we know that $\boldsymbol{Q}(^3/_2)$ is non-empty and an open subset of $(\mathbb{R}^+)^4 x\mathbb{R}$. Then the foliated scalar-scalar multiverse determined by $\boldsymbol{Q}(^3/_2)$ would be an 9-dimensional space, FMV($\boldsymbol{Q}(^3/_2)$):= $\boldsymbol{Q}(^3/_2)$x $\mathbb{R}^4$, with coordinates: (($\alpha$, $\beta$, $\gamma$, $k_1$,$k_2$), (t,u,v,w)). For the model described above we choose a point ($\alpha$, $\beta$, $\gamma$, $k_1$, 0) in $\boldsymbol{Q}(^3/_2)$, and attach to that point a universe whose metric is:

(i)   degenerate for t<0, $ds^2 = -dt^2 + 0 \cdot (du^2 + dv^2 + dw^2)$;

(ii)  determined by $\varphi'_{\gamma,k1,0}$ for $0 \leq t \leq t_1$: and

(iii) determined by $\varphi'_{\alpha,\beta}$ for $t_1 < t$.

It should be noted that in general at the times $t_1$ and $t_2$ when $\varphi'_{\alpha,\beta} = \varphi'_{\gamma,k1,0}$, we have $\varphi_{\alpha,\beta} \neq \varphi_{\gamma,k1,0}$. So that if we try to construct one scalar field $\varphi$ for all time by

$$\varphi := \begin{cases} 0, & \text{for } t \leq 0, \\ \varphi_{\gamma,k1,0} & \text{for } 0 < t \leq t_1, \\ \varphi_{\alpha,\beta} & \text{for } t_1 < t, \end{cases}$$

then this $\varphi$ is discontinuous at $t=t_1$.  However, $\varphi'$ does admit a continuous extension through $t=t_1$, if we define

$$\varphi'(t_1) := \lim_{t \to t_1^-} \varphi'_{\gamma,k1,0}(t) = \lim_{t \to t_1^+} \varphi'_{\alpha,\beta}(t) \ .$$

$\varphi$, with $\varphi'$ as just defined, is an element of a class of functions that I call class $C^{\infty,1}$.  More generally I define this class of functions in the following way.  Let $\psi: \mathbb{R} \to \mathbb{R}$ be defined on an interval of the form $(a,b) \subset \mathbb{R}$ ( where a could be $-\infty$ and b could be $+\infty$), and let $\{P_i\}_{i \in I}$ be distinct points in $(a,b)$, where $I \subset \mathbf{Z}$ (the set of integers) is a non-empty indexing set, which could be countably infinite.  I shall assume that the set $\{P_i\}_{i \in I}$ does not have a limit point, and thus $D := (a,b) \backslash \{P_i\}_{i \in I}$ is an open subset of $\mathbb{R}$. $\psi$ is said to be of class $C^{k,1}$ on $(a,b)$ ($k \in \mathbb{N}$, the set of natural numbers) if:

(i)  $\psi$ is discontinuous at the points $\{P_i\}_{i \in I}$,

(ii) $\psi$ is of class $C^k$ on $D$, and

(iii) $\forall \ P_i$, with $i \in I$, $\lim\limits_{t \to P_i^-} \psi'(t) = \lim\limits_{t \to P_i^+} \psi'(t) =: \psi'(P_i)$ .

Thus when a function is of class $C^{k,1}$ on $(a,b)$ it has a well-defined continuous "first derivative" on $(a,b)$, even though it is discontinuous at points in $(a,b)$.

When building cosmological spaces from the solutions obtained in **Section 3**, the scalar field $\varphi$ we obtain by piecing solutions together when their first derivatives cross, will be of class $C^{\infty,1}$, which is adequate for our needs since the metric tensor $g_{ij}$ depends upon $\varphi'$ and not $\varphi$.  Thus $g_{ij}$ will be continuous, but its derivatives can experience discontinuities at those points where φ jumps between solutions. In what follows whenever I talk about a physical quantity, such as the scalar curvature R, being continuous across a point where φ experiences a discontinuity, it will be in the sense that its left and right limits are equal at those points.

Let $\boldsymbol{Q'}(^3/_2) := \{(\alpha,\beta,\gamma) \mid (\alpha,\beta,\gamma,1,0)\in\boldsymbol{Q}(^3/_2)\}$. $\boldsymbol{Q'}(^3/_2)$ is an open subset of $(\mathbb{R}^+)^3$. (Note that by working with $\boldsymbol{Q'}(^3/_2)$ what I have done is pulled out the constant $k_1$ appearing in $\varphi = \gamma(k_1 t + k_2)^{3/2}$, absorbing it into γ, and set $k_2=0$, so that our solutions start at t=0, with φ'(0)=0.) In Appendix B, I describe one way in which the 4-dimensional leaves of the disconnected 7-dimensional manifold FMV($\boldsymbol{Q'}(^3/_2)$) can be "bound" together to give rise to a connected 4-dimensional manifold. I shall now describe a second way of accomplishing this using quotient manifolds.

To that end, if $(\alpha,\beta,\gamma) \in\boldsymbol{Q'}(^3/_2)$, let

$L_{(\alpha,\beta,\gamma)} := \{ ((\alpha,\beta,\gamma), (t,u,v,w)) \mid (t,u,v,w) \in\mathbb{R}^4 \}$.

FMV($\boldsymbol{Q'}(^3/_2)$) is foliated by the leaves $L_{(\alpha,\beta,\gamma)}$. Each leaf has a metric tensor defined by:

(i)   $ds^2 = -\,dt^2 + 0(du^2 + dv^2 + dw^2)$ , t<0;

(ii)  $ds^2 = -\,dt^2 + [^3/_2\gamma t^{1/2}]^2(du^2 + dv^2 + dw^2)$, $0\underline{\le} t < t_1$; and

(iii) $ds^2 = -\,dt^2 + [\alpha\beta e^{\beta t}]^2(du^2 + dv^2 + dw^2)$, $t_1\underline{\le} t$ .

Evidently, the open subsets of the leaves $L_{(\alpha,\beta,\gamma)}$ where t<0, have the same degenerate geometry and are all diffeomorphic to $\{(t,u,v,w)\in\mathbb{R}^4 \mid t<0\}$. This observation can be used to "glue" these leaves together in the following manner. Let us define an equivalence relation ~ on FMV($Q'(^3/_2)$) by:

(i)     $((\alpha_1,\beta_1,\gamma_1),\ (t_1,u_1,v_1,w_1)) \sim ((\alpha_2,\beta_2,\gamma_2),\ (t_2,u_2,v_2,w_2))$ if $t_1=t_2<0$, and $u_1=u_2$, $v_1=v_2$ $w_1=w_2$; and

(ii)    $((\alpha_1,\beta_1,\gamma_1),\ (t_1,u_1,v_1,w_1)) \sim ((\alpha_2,\beta_2,\gamma_2),\ (t_2,u_2,v_2,w_2))$ if $\alpha_1=\alpha_2$, $\beta_1=\beta_2$, $\gamma_1=\gamma_2$, $t_1=t_2\geq0$, $u_1=u_2, v_1=v_2$, $w_1=w_2$.

The resulting quotient set, FMV($\mathbf{Q'}(^3/_2)$)/~, can be endowed with a $C^\infty$ structure of dimension 4 which is such that the natural surjection $\mu$: FMV($\mathbf{Q'}(^3/_2)$)→FMV($\mathbf{Q'}(^3/_2)$)/~ is a submersion. In Appendix B, I put this $C^\infty$ structure on FMV($Q'(^3/_2)$)/~. More exactly, I put the $C^\infty$ structure on a connected 4-dimensional space, MV($Q'(^3/_2)$) which is bijectively related to FMV($Q'(^3/_2)$)/~. This bijection allows us to put a $C^\infty$ structure of dimension 4 on FMV($Q'(^3/_2)$)/~, turning it into a quotient manifold diffeomorphic to MV($Q'(^3/_2)$).

I shall now describe two ways in which we can get a universe in the multiverse to end. Suppose that $(\alpha, \beta, \gamma) \in Q'(^3/_2)$, and that $t = t_1$ and $t = t_2$, with $0<t_1<t_2$, are the times when $\varphi'_{\alpha,\beta} = \varphi'_{\gamma,1,0}$. We can choose a negative number $\acute{\alpha}$ so that $\varphi'_{\acute{\alpha},-\beta}(t) := \acute{\alpha}e^{-\beta t}$ is a solution to $H_{2,4}=0$, with $\varphi'_{\acute{\alpha},-\beta}>0$, and $\varphi'_{\acute{\alpha},-\beta}(t_2) = \varphi'_{\alpha,\beta}(t_2)$. (It is easily seen that $\acute{\alpha}=-\alpha\exp(2\beta t_2)$.) So as time, t, moves through $t_2$, $\varphi$ can jump from $\varphi_{\alpha,\beta}(t)$ to $\varphi_{\acute{\alpha},-\beta}(t)$ in such a way that the metric tensor is continuous, but its first derivatives experience a discontinuity, with R being continuous as we pass through $t_2$, since R = $12\beta^2$ (see **Figure 1** on page 81, where the solid blue curve represents the model universe's history). In this model, the originally expanding universe, gradually collapses down to the $\varphi' = 0$ state, over an infinitely long period. Note that in this model the three branches of the universe (determined by $\varphi'$) meet when t = $t_2$ (although the first branch was left behind as t increased past $t_1$). Of course we could have chosen other exponentially decreasing solutions of $H_{2,4}$=0, to end our model universe, but for all of these other solutions R would experience a discontinuity as t passes through $t_2$.

Next, let us construct a model that ends more abruptly in a finite amount of time.

Let's consider the three solutions to $H_{2,4} = 0$, given by

$$\varphi_{1,1} := \gamma_1(t)^{3/2}, \quad \varphi_{1,2} := \alpha_1 \exp(\beta_1 t), \quad \varphi_{1,3} := -\gamma_1(2t_2 - t)^{3/2} , \qquad \text{Eq.4.2}$$

where $(\alpha_1, \beta_1, \gamma_1) \in \boldsymbol{Q'}(^3/_2)$. The first two solutions in Eq.4.2 are defined for t$\geq$0, while the last solution is defined for t$\leq$2$t_2$. In addition we have $\varphi'_{1,1}(t_2) = \varphi'_{1,2}(t_2) = \varphi'_{1,3}(t_2)$ . This triple of solutions determines a universe which begins explosively with $\varphi$ in the 1,1 state. Then $\varphi$ jumps at t = $t_1$ into the 1,2 state, and then jumps at t = $t_2$ into the 1,3 state, where it continues until t = 2$t_2$, and the universe comes to an abrupt end (see **Figure 2**, on page 81). It should be noted that in this model the graph of the third leg of the universe is actually the reflection of the graph of the first leg of the universe in the line t=$t_2$, if that first leg is continued to time t=$t_2$. Evidently this universe need not end at time $t = 2t_2$, because we can find other solutions of the form $\varphi = \gamma_2(t + k_2)^{3/2}$ which begin explosively at a time after $2t_2$. In a moment I shall construct a series of such models, but before I do, I would like to make one remark.

In the model just constructed, R jumps from R=0 to R=$12\beta^2$ as we pass through $t=t_1$, and then jumps from R=$12\beta^2$ back to R=0, as we pass through $t=t_2$. It makes one think that the universe had to absorb some scalar curvature to jump from one $\varphi$ state to the next, and then had to emit that scalar curvature to return to the "ground state." This is reminiscent of the electron in the hydrogen atom absorbing a photon to move from the ground state to a higher energy state, and then emitting that photon to return to the ground state. But in this analogy, where did the universe borrow the scalar curvature from, and then return it? The obvious place is from the R=0 vacuum states. Perhaps the first R=0 vacuum state solution, $\varphi_{1,1}$, "creates" an amount of positive scalar curvature R which enables the second vacuum state, $\varphi_{1,2}$, to come into existence with $\beta_1 = (R/12)^{\frac{1}{2}}$. After a period of time this second vacuum state must return this scalar curvature, which it then "gives" to the third, R=0, vacuum state $\varphi_{1,3}$. Of course, this is just speculation on my part. What someone needs to do at this juncture is quantize the scalar field $\varphi$, and in the process the gravitational field $g_{ij}$, to see if that lends any credence to the above thoughts. Evidently the canonical quantization approach is not available, since the Lagrangians $L_4$, $L_5$ and $L_C$ that we have employed, have Ostrogradsky instability issues. Hence we shall require a new approach to quantize $\varphi$. Perhaps a Feynmann path integral approach might work, but that also has difficulties when the Lagrangian is second, or third-order. Hopefully this new approach will give rise to an inequality relating the lifetime, $T_\ell := t_2 - t_1$, and $\beta$ in the exponential solution. I would like to see something like $T_\ell \beta < q = {}^3/_2$, or $T_\ell \beta < q - 1 = \frac{1}{2}$.

Now for more models.

$\forall$ integer $n \in \mathbb{N}$, let $t_{nb} > 0$, denote the time that our $n^{th}$ universe begins, with $t_{1b} = 0$. Following Eq.4.2, the $n^{th}$ universe will be built from three solutions of $H_{2,4} = 0$ defined by:

$$\varphi_{n,1} := \gamma_n (t - t_{nb})^{3/2}, \, t_{nb} \leq t < t_{n1,}$$

$$\varphi_{n,2} := \alpha_n \exp(\beta_n t), \, t_{n1} \leq t \leq t_{n2} \, , \text{ and} \qquad \qquad \text{Eq.4.3}$$

$$\varphi_{n,3} := -\gamma_n (2t_{n2} - t_{nb} - t)^{3/2}, \, t_{n2} < t \leq 2t_{n2} - t_{nb}$$

where $(\alpha_n, \beta_n, \gamma_n, 1, -t_{nb}) \in \boldsymbol{Q}(^3/_2)$, with $t_{n1} < t_{n2}$ being the times when $\varphi'_{n,1}$ and $\varphi'_{n,2}$ meet. The time, $t_{ne}$, when this universe ends is $t_{ne} = 2t_{n2} - t_{nb}$, and the length of its duration is $t_{ne} - t_{nb} = 2(t_{n2} - t_{nb})$. It will be assumed that $\forall \, n \in \mathbb{N}$, the constants $\alpha_n$, $\beta_n$, $\gamma_n$, and $t_{nb}$ will have been chosen so that $t_{ne} \leq t_{(n+1)b}$.

It should now be clear that by stringing the above countably infinite collection of universes together, we have constructed a Universe comprised of an endless chain of "mini-universes," each of which begins explosively (with the first beginning at $\varphi' = 0$, when t=0), then ends abruptly after a decline, with a period of exponential growth between these two extreme states, which can be made as flat as you wish (see **Figure 3**, on page 82). One should note that the mini-universes given by Eq.4.3 spend as much time expanding as they do collapsing.  In addition, if $\forall\ n\in\mathbb{N}$, $t_{ne}=t_{(n+1)b}$, then our mini-universes bounce from one universe into the next without any hesitation.

The models presented above, which were based upon solutions of $H_{2,4} = 0$, could be made to end more abruptly if we wished.  We could also extend the infinite chain of models presented in **Figure 3** so that they extend back into the infinite past to obtain a Universe without beginning or end. This Universe would  consist of a countably infinite series of mini-universes, each of  which exist for only a finite amount of time from their beginning to their end. However, each of these mini-universes begin and end with a curvature singularity where $R^{abcd}R_{abcd}$ blows up.  These singularites can be avoided if we made the mini-universes overlap a bit so that each begins and ends with $\varphi'>0$. (Perhaps when $\varphi'$ x 1cm equals the Planck length.) By doing this we would also avoid the curvature singularity when $\varphi' = 0$, but we would still have a discontinuity in the metric tensor at the moment when one universe ends, and the other begins.

In passing I would like to point out that  we could also build a solution to $H_{2,4}=0$, just using exponential solutions $\varphi_{\alpha,\beta} = \alpha e^{\beta t}$, which is even more in keeping with the usual inflationary model. For $\varphi_{\alpha,\beta} = \alpha e^{\beta t}$, we have $\varphi'_{\alpha,\beta}=\alpha\beta e^{\beta t}$, and hence $\varphi'_{\alpha,\beta}(0)=\alpha\beta$.  Thus for the non-constant exponential solutions we always have $\varphi'_{\alpha,\beta}$ beginning at a non-zero value when time begins at t=0. But we can suppose that for t≤0; *i.e.,* prior to the expansion of the universe, the universe did not satisfy any particular equations and just sat there with the metric on the t=constant slices being $(\alpha\beta)^2(du^2+dv^2+dw^2)$. Consequently when t≤0 the universe had plenty of time to achieve thermodynamic equalibrium, in fact an infinite amount of time, if that was necessary, prior to the "awakening" of $\varphi_{\alpha,\beta}$ at t=0. As time advances through t=0 the universe described by $\varphi_{\alpha,\beta}$ could jump to another "very flat" exponential solution, $\varphi_{\alpha',\beta'}$ with β' very close to zero. As time continues, this exponential solution could jump to another, say, decreasing exponential solution $\varphi_{\alpha'',\beta''}$ (*see,* **Figure 4**, on page 83). We could continue stringing exponential solutions together, but any such solution would describe a universe which continues forever due to the nature of our exponential solutions. In addition the exponential solutions are free of curvature singularities, but the differentiability of the metric breaks down where the different exponential solutions join.

Everything we did with the solutions to $H_{2,4} = 0$, with q = $^3/_2$, could be done with our solutions to $H_{2,5} = 0$ and $H_C = 0$, for q= $^3/_2$ as well as other values of q. However, when q≠$^3/_2$, Eq.3.11 tells us that the value of R for the solutions of the form $\gamma(k_1t+k_2)^q$, is not constant, but is positive if $^3/_2$<q<2, while R is a positive constant for each solution of the form $\alpha e^{\beta t}$. So when $^3/_2$<q<2 a solution of the form $\gamma(k_1t+k_2)^q$ (which begins explosively when t= −$k_2$/$k_1$) can jump to a solution of the form $\alpha e^{\beta t}$ when their R values are equal, provided α, β, γ, $k_1$ and $k_2$ are chosen so that $g_{ij}$ is continuous across the jump.

All of the examples of scalar-scalar theories I have presented in this paper are based upon the FLRW cofinsler function f given in Eq.2.7, which is intended to be used for cosmological considerations. Other choices can be made for f to deal with different situations in the universe. *E.g.,* to deal with a few astronomical bodies we could choose the trivial f; *viz.,*

$$f := g^{ij} \circ \pi \, y_i \, y_j \, . \qquad\qquad Eq.4.4$$

In this case when we build the metric tensor using Eq.2.4, with φ chosen equal to a constant, we obtain a Lorentzian metric. When this metric is used in the Horndeski Lagrangian presented in Eq.3.4, with φ constant, we obtain the Einstein Lagrangian with cosmological term. This shows how we can recover Einstein's theory from a scalar-scalar field theory using the Horndeski Lagrangian. Being able to do this is fairly important due to the success of Einstein's theory in describing situations with only a handful of astronomical bodies.

The next biggest region of astronomical interest would be galaxies and galactic clusters. I am not certain what choice of f would be suitable for this situation. If we simply chose the trivial f given in Eq. 4.4 above, with φ not a constant, then the theory we would arrive at would be the usual Horndeski theory.

This completes my introduction to scalar-scalar field theories. I hope that you have found it interesting.

**Acknowledgements**

I would like to thank Dr. M. Zumalaćarregui, Dr. J.M. Ezquiagua and Dr. K. Takahashi for discussions on some of the topics dealt with in this paper. I owe a debt of gratitude to Dr. A. Guarnizo Trilleras for allowing me to read his Ph.D. thesis, which introduced me to the subject of beyond Horndeski theory, and got me back into working on scalar-tensor theories.  Professor A. Silvestri, and the Lorentz Institute of Physics at Leiden University in the Netherlands, deserves my thanks for their financial support through the acquisition of one of my paintings dealing with Horndeski Theory for their Institution.  Lastly, I need to thank Dr. Moon Nahm, M.D., for not only encouraging me to pursue the research I have presented here, but also for financial assistance through the purchase of a painting of mine for the Physics Department at Washington University in St. Louis, where I did my undergraduate studies.

**Appendix A:  Vacuum Solutions for Various  Hamiltonians and More**

In this appendix I shall discuss the solutions to $H_{2,4}$=0, $H_{2,5}$=0 and $H_C$=0, when these Hamiltonians are given by Eqs. 3.79, 3.80 and 3.81. I shall begin with the equation, $H_{2,4}=0$, since it is the only equation of the three for which all solutions can be explicitly found and expressed in terms of elementary functions. After completing these tasks I shall determine all of the solutions to $H$ = constant $\neq 0$, where $H$ is given by Eq.3.82.

Recall that we obtained the Hamiltonian $H_{2,4}$ presented in Eq.3.80 by demanding that the original $H_{2,4}$ given in Eq.3.35 admit vacuum solutions of the form $\varphi=\alpha e^{\beta t}$ and $\varphi=\gamma(k_1 t+k_2)^{3/2}$. However, once these conditions have been imposed, the resulting Hamiltonian of Eq.3.80 may have additional vacuum solutions. That is what I want to determine. After some simplification, the equation of interest to us becomes

$$2(\varphi')^3 - 5\varphi\varphi'\varphi'' + 3\varphi^2\varphi''' = 0. \qquad\qquad \text{Eq.A.1}$$

Since Eq.A.1 is a third-order ordinary differential equation we know that the general solution will have three constants of integration, while there are really only two integration constants in the aforementioned two solutions of this equation. Hence we suspect that Eq.A.1 has other solutions for us to discover.

We begin our search for solutions to Eq.A.1 by finding an integrating factor. To that end we multiply Eq.A.1 by $\varphi^b$ and try to choose a value of b so that the resulting equation is exact. In this way we obtain two integrating factors, $\varphi^{-3}$ and $\varphi^{-5/3}$. Using the first integrating factor we see that Eq.A.1 is equivalent to

$$\underline{d} \, (3\varphi^{-1}\varphi'' - \varphi^{-2}(\varphi')^2) = 0 \ ,$$
$$dt$$

and hence

$$3\varphi^{-1}\varphi'' - \varphi^{-2}(\varphi')^2 = 3\sigma \ , \qquad\qquad\qquad Eq.A.2$$

where $\sigma$ is a constant. To solve this equation we make the substitution

$$\psi := \varphi^{-1}\varphi' \ . \qquad\qquad\qquad Eq.A.3$$

Upon doing so, Eq.A.2 becomes the autonomous first order differential equation

$$\psi' + {}^2\!/_3\,\psi^2 = \sigma \ . \qquad\qquad\qquad Eq.A.4$$

When $\sigma$ is non-negative, a trivial solution to this equation is $\psi = \pm({}^3\!/_2\sigma)^{1/2}$ . Using this in Eq.A.3 gives us

$$\varphi = \alpha \exp(\pm({}^3\!/_2\sigma)^{1/2}t), \qquad\qquad\qquad Eq.A.5$$

where $\alpha$ is an integration constant. This is one of our previous solutions to $H_{2,4} = 0$.

We shall now seek other non-trivial solutions to Eq.A.4. To that end let us rewrite this equation as

$$\frac{d\psi}{\sigma - {}^2\!/_3\,\psi^2} = dt,$$

and hence

$$\int(\sigma - {}^2\!/_3\,\psi^2)^{-1}d\psi = t + t_0 \ , \qquad\qquad\qquad Eq.A.6$$

where $t_0$ is an integration constant. To solve this equation there are three cases to consider; *viz.,* $\sigma = 0$, $\sigma > 0$ and $\sigma < 0$. For the latter two cases we shall set $\sigma = {}^3\!/_2\tau^2$ and $\sigma = -{}^3\!/_2\tau^2$ respectively, where $\tau$ is a non-zero constant. We shall examine each of these cases in turn.

Case (i): $\sigma = 0$.  Using this choice of $\sigma$ in Eq.A.6 gives us $\psi = {}^3/_2 (t + t_0)^{-1}$.  Thus we can use Eq.A.3 to conclude that

$$\varphi = \gamma |t + t_0|^{3/2}, \qquad\qquad\qquad\qquad \text{Eq.A.7}$$

where $\gamma$ is a constant.  This is the second of our two known solutions to $H_{2,4} = 0$.

Case (ii): $\sigma = {}^3/_2 \tau^2$.  Eq.A.6 can be rewritten as

$$3\tau(t + t_0) = \int (\tau - {}^2/_3 \psi)^{-1} d\psi + \int (\tau + {}^2/_3 \psi)^{-1} d\psi. \qquad\qquad \text{Eq.A.8}$$

Upon performing the integrals in Eq.A.8 we find that

$$2\tau(t + t_0) = \ln(|(\tau + {}^2/_3 \psi)(\tau - {}^2/_3 \psi)^{-1}|) ,$$

which  can be solved for $\psi$, to yield

$$\psi = {}^3/_2 \tau \tanh(\tau(t + t_0)) .$$

Upon combining this equation with Eq.A.3 we discover that

$$\varphi = \Omega \cosh^{3/2}(\tau(t + t_0)) , \qquad\qquad\qquad \text{E q.A.9}$$

where $\Omega$ is an integration constant.

Case (iii) $\sigma = -{}^3/_2 \tau^2$.   For this case Eq.A.6 becomes

$${}^{-3}/_2 (t + t_0) = \int (\tau^2 + ({}^2/_3)^2 \psi^2)^{-1} d\psi ,$$

and hence

$$\psi = -{}^3/_2 \tau \tan(\tau(t + t_0)) .$$

This permits us to employ Eq.A.3 to deduce that

$$\varphi = \Omega \, |\cos(\tau(t + t_0))|^{3/2} \, , \qquad\qquad\qquad Eq.A.10$$

where $\Omega$ is an integration constant.

Let us now examine whether the second integrating factor of Eq.A.1; *viz.,* $\varphi^{-5/3}$, leads to other solutions of $H_{2,4}=0$. Upon multiplying Eq.A.1 by $\varphi^{-5/3}$ we obtain an exact differential equation, which can be integrated to give us

$$\varphi^{1/3}\varphi'' - \varphi^{-2/3}(\varphi')^2 = \sigma \, , \qquad\qquad\qquad Eq.A.11$$

where $\sigma$ is an integration constant. If $\sigma=0$, then the solution to Eq.A.11 has the familiar form $\varphi=\alpha e^{\beta t}$. So let us now assume that $\sigma \neq 0$. Then $\varphi^{-7/3}\varphi'$ is an integrating factor for this equation. When Eq.A.11 is multiplied by this factor the resulting equation implies that

$$(\varphi')^2 = \lambda\varphi^2 - {}^3\!/_2 \, \sigma\varphi^{2/3} \, , \qquad\qquad\qquad Eq.A.12$$

and hence

$$\varphi' = \varphi^{1/3}|\lambda\varphi^{4/3} - {}^3\!/_2\sigma|^{1/2}$$

where $\lambda$ is an integration constant. This in turn implies that

$$\int \varphi^{-1/3}|\lambda\varphi^{4/3} - {}^3\!/_2\sigma|^{-1/2} \, d\varphi = (t-t_0).$$

This integral can be computed making the substitution $\psi:=\varphi^{2/3}$. Upon examining the various integrals so obtained we get the solutions to Eq.A.12 to be

$$\varphi = \Omega \, |\sinh(\tau(t + t_0))|^{3/2} \qquad\qquad\qquad Eq.A.13$$

and

$$\varphi = \Omega \, |\sin(\tau(t + t_0))|^{3/2}, \qquad\qquad\qquad\qquad \text{Eq.A.14}$$

along with the solutions given by Eqs. A.9 and A.10, when the values of $\Omega$ and $\tau$ are chosen suitably.

Of course the sine and cosine solutions given in Eqs.A.10 and A.14 are essentially the same solution because of our freedom to chose the phase factor $t_0$. However, in practice it is useful to regard them as different when looking for solutions which begin with $\varphi'(0) = 0$, and $t_0 = 0$.

Eqs. A.5, A.7, A.9, A.10, A.13 and A.14 provide us with all of the solutions to the equation $H_{2,4} = 0$, where $H_{2,4}$ is given by Eq.3.80. The new solutions are those involving cosh, sinh, cosine and sine, each of which involve three constants of integration. All of these new classes of solutions yield $\varphi'=0$, at $t=0$, when we choose $t_0=0$. However, the cosh and cosine solutions do not begin explosively (*i.e.,* with a vertical tangent vector for $\varphi'$) at $t=0$, and yet R blows up at $t=0$ for these solutions. While $\varphi'$ for the sinh and sine solutions begins explosively at $t=0$, with R being finite there. Nevertheless all these new solutions lead to curvature singularities at $t=0$, where

$$R^{abcd}R_{abcd} = 12((\varphi')^{-2}(\varphi''')^2 + (\varphi')^{-4}(\varphi'')^4)$$

blows up. The cosh and sinh solutions presented in Eqs.A.9 and A.13, provide us with an exponential like solution that yields $\varphi'=0$ at $t=0$ (when $t_0=0$).

Let us now turn our attention to the solution of $H_{2,5} = 0$, where $H_{2,5}$ is given by Eq.3.81. Due to Eq.3.78 we see that if we can find all of the solutions to $H_{2,5} = 0$, then we shall also have found all of the solutions to $H_C = 0$.

Using Eq.3.81 we find that when the equation $H_{2,5} = 0$ is multiplied by $-\frac{1}{12}\varphi^3(\varphi')^{-3}$ we obtain

$$\varphi^{-2}(\varphi')^4 - 3\varphi^{-1}(\varphi')^2\varphi'' + (\varphi'')^2 + \varphi'\varphi''' = 0,$$

or equivalently

$$\frac{d}{dt}(\varphi'\varphi'' - \varphi^{-1}(\varphi')^3) = 0 \ .$$

Hence

$$\varphi'\varphi'' - \varphi^{-1}(\varphi')^3 = -\frac{2}{3}\sigma \ , \qquad\qquad\qquad Eq.A.15$$

where $\sigma$ is an integration constant. When $\sigma = 0$ the solution to Eq.A.15 is the well-known solution $\varphi = \alpha e^{\beta t}$. So we shall now assume that $\sigma \neq 0$ in Eq.A.15. For that case an integrating factor for this differential equation is $\varphi^{-3}\varphi'$. When Eq.A.15 is multiplied by this factor the resulting equation can be rewritten as follows:

$$\varphi^{-3}(\varphi')^3 - \sigma\varphi^{-2} = \lambda \ , \qquad\qquad\qquad Eq.A.16$$

where $\lambda$ is an integration constant. If $\lambda = 0$, then the solution to Eq.A.16 can be written as $\varphi = \gamma|t + t_0|^{3/2}$, which is another of our known solutions to $H_{2,5} = 0$. Thus we shall now assume that $\lambda \neq 0$ in Eq.A.16. The solution to the resulting autonomous differential equation can be solved by evaluating

$$\int \varphi^{-1/3}(\sigma + \lambda\varphi^2)^{-1/3}d\varphi = (t + t_0) \ . \qquad\qquad\qquad Eq.A.17$$

To assist in computing the above integral we make the substitution, $\psi := \varphi^{2/3}$, to find that Eq.A.17 becomes

$$\int (\kappa + \lambda \psi^3)^{-1/3} d\psi = \tfrac{2}{3}(t + t_0) \,. \qquad\qquad Eq.A.18$$

Unfortunately I cannot evaluate this integral in terms of elementary functions. Nevertheless, we see that the equation $H_{2,5} = 0$, and hence $H_C = 0$, will have some solutions that are different than those for $H_{2,4} = 0$.

Let's turn our attention to the Hamiltonian $H$ given in Eq.3.82. I mentioned earlier that all the solutions to $H = 0$ are given by $\varphi = \alpha e^{\beta t}$ and $\varphi = \gamma (k_1 t + k_2)^q$, where $\alpha$, $\beta$, $\gamma$, $k_1$, $k_2$ and $q$ are arbitrary, except for the requirement that $\varphi' > 0$. I shall now determine all of the non-vacuum solutions to $H = \text{constant} = 2\kappa\kappa_1$. To that end we write this equation as

$$\frac{d}{dt} (\varphi(\varphi')^{-2}\varphi'') = -2\kappa_1 \,, \qquad\qquad Eq.A.19$$

where $\kappa_1 \neq 0$. This equation implies that

$$\varphi(\varphi')^{-2}\varphi'' = -(2\kappa_1 t + \kappa_2 - 1),$$

where $-(\kappa_2 - 1)$ is an integration constant. Thus $\varphi$ must satisfy

$$\frac{\varphi''}{\varphi'} = \frac{-\varphi'}{\varphi} (2\kappa_1 t + \kappa_2 - 1) \,. \qquad\qquad Eq.A.20$$

To solve this equation we make the substitution

$$\xi := \varphi^{-1}\varphi' \qquad\qquad Eq.A.21$$

and hence

$$(\varphi')^{-1}\varphi'' = \xi^{-1}\xi' + \xi \ . \qquad\qquad\qquad\qquad \text{E q.A.22}$$

Eqs. A.20-A.22 imply that $\xi$ must satisfy

$$\xi^{-2}\xi' = -(2\kappa_1 t + \kappa_2)$$

and hence

$$\xi = (\kappa_1 t^2 + \kappa_2 t + \kappa_3)^{-1}, \qquad\qquad\qquad\qquad \text{Eq.A.23}$$

where $\kappa_3$ is an integration constant. When this is combined with Eq.A.21 we discover

that

$$\varphi = A\exp[\int(\kappa_1 t^2 + \kappa_2 t + \kappa_3)^{-1}dt] \ , \qquad\qquad\qquad \text{Eq.A.24}$$

where A is the third, and final, integration constant of our third order differential equation. In order to evaluate the integral appearing in Eq.A.24 there are three cases to consider: $\delta=0$, $\delta>0$ and $\delta<0$, where $\delta:= 4\kappa_1\kappa_3 - \kappa_2^2$. We shall examine each case in turn.

Case (i): $\delta = 0$. In this case we can write $\kappa_1 t^2 + \kappa_2 t + \kappa_3 = \kappa_1^{-1}(\kappa_1 t + \tfrac{1}{2}\kappa_2)^2$. Using this

in Eq.A.24 gives us

$$\varphi = A \exp(-(\kappa_1 t + \tfrac{1}{2}\kappa_2)^{-1}), \qquad\qquad\qquad \text{Eq.A.25}$$

and thus

$$\varphi' = \kappa_1(\kappa_1 t + \tfrac{1}{2}\kappa_2)^{-2}\varphi \ , \qquad\qquad\qquad\qquad \text{Eq.A.26}$$

with both $\varphi$ and $\varphi'$ being defined on $\mathbb{R} \setminus \{-\kappa_2/(2\kappa_1)\}$. Since we are requiring $\varphi' > 0$,

Eqs.A.25 and A.26 imply that $A\kappa_1 > 0$. So when $\varphi>0$, $\kappa_1$ must be positive, and hence

$H$ = constant > 0 (assuming $\kappa > 0$). Similarly when $\varphi < 0$, $\kappa_1$ must be negative, and hence $H$= constant <0.

Case (ii): $\delta > 0$. Using Eq.A.24 we find that

$$\varphi = A\exp[2\delta^{-\frac{1}{2}}\tan^{-1}(\delta^{-\frac{1}{2}}(2\kappa_1 t + \kappa_2))] \,, \qquad\qquad Eq.A.27$$

and hence

$$\varphi' = 4\kappa_1(\delta + (2\kappa_1 t + \kappa_2)^2)^{-1}\,\varphi \,, \qquad\qquad Eq.A.28$$

where $\varphi$ and $\varphi'$ are well-defined for all t. Thus we see that for $\varphi'$ to be positive, we once again require $A\kappa_1 > 0$. So in this case we have the same relationship between the signs of $\varphi$ and $H$ as we did in the above case.

Case (iii) $\delta < 0$. In this case $\kappa_1 t^2 + \kappa_2 t + \kappa_3$ can be factored, and the integral in Eq.A.24 gives us

$$\varphi = A\,|\{(2\kappa_1 t + \kappa_2 - (-\delta)^{\frac{1}{2}}) / (2\kappa_1 t + \kappa_2 + (-\delta))\}|^{1/\sqrt{-\delta}} \,. \qquad\qquad Eq.A.29$$

Employing Eqs.A.21 and A.23 we find that

$$\varphi' = 4\kappa_1[(2\kappa_1 t + \kappa_2)^2 + \delta]^{-1}\,\varphi \,. \qquad\qquad Eq.A.30$$

Since $\delta < 0$ we can not use Eqs.A.29 and A.30 to simply say that if $A\kappa_1 > 0$ then $\varphi' > 0$. So to determine the relationship between the sign of $\kappa_1$ and $H$ we proceed as follows.

When $\delta \equiv 4\kappa_1\kappa_3 - \kappa_2^2 < 0$, the equation $\xi^{-1}= \kappa_1 t^2 + \kappa_2 t + \kappa_3 = 0$, has two distinct zeroes that we shall denote by $z_L$ and $z_R$ where $z_L < z_R$. $\varphi$ is undefined at the zero given by $-((\kappa_2 + (-\delta)^{1/2})/(2\kappa_1))$, while $\varphi' = \xi\varphi$ is undefined at both $z_L$ and $z_R$. Hence the domain of $\varphi$ is comprised of the three pieces: $t < z_L$, $z_L < t < z_R$, and $z_R < t$. Let $\varphi_L$ denote the restriction of $\varphi$ to $t < z_L$, $\varphi_M$ denote the restriction of $\varphi$ to $z_L < t < z_R$, and $\varphi_R$ denote the restriction of $\varphi$ to $z_R < t$.

If $\kappa_1 > 0$, then $\xi > 0$ for $t < z_L$ or $z_R < t$. Therefore since $\varphi' = \xi\varphi$, $\varphi_L$ and $\varphi'_L$ have the same sign, as do $\varphi_R$ and $\varphi_R'$. While $\varphi_M$ and $\varphi_M'$ have opposite signs.

If $\kappa_1 < 0$, then $\xi < 0$ for $t < z_L$ or $z_R < t$. Therefore $\varphi_L$ and $\varphi_L'$ have opposite signs, as do $\varphi_R$ and $\varphi_R'$. While $\varphi_M$ and $\varphi_M'$ have the same sign.

This leads us to conclude that when $\delta < 0$:

(i)  If $\kappa_1 > 0$, and we want $\varphi > 0$, along with $\varphi' > 0$, then we must work with $\varphi_L$ and $\varphi_R$, and $A\kappa_1 > 0$.

(ii)  If $\kappa_1 > 0$, and we want $\varphi < 0$, along with $\varphi' > 0$, then we must work with $\varphi_M$, and $A\kappa_1 < 0$.

(iii)  If $\kappa_1 < 0$, and we want $\varphi > 0$, along with $\varphi' > 0$, then we must work with $\varphi_M$, and $A\kappa_1 < 0$. And lastly,

(iv)  If $\kappa_1 < 0$, and we want $\varphi < 0$, along with $\varphi' > 0$, then we must work with $\varphi_L$ and $\varphi_R$, and $A\kappa_1 > 0$.

In the previous two cases, where $\delta = 0$, or $\delta > 0$, we found that the sign of $\varphi$ was the same as the sign of the Hamiltonian $H$. For the case, where $\delta < 0$, we see that the sign of $\varphi_L$ and $\varphi_R$ are the same as the sign of $H$, while the sign of $\varphi_M$ is the opposite of the sign of $H$ (recall that we are assuming that $\kappa > 0$). Hence if we wanted the maximum connected domain of our solutions to $H = 2\kappa\kappa_1$ to be unbounded, then the sign of $H$ and $\varphi$ would be the same.

Let us now consider the behavior of R for the above three solutions to $H = 2\kappa\kappa_1$. From Eq.3.11 we know that

$$R = 6((\varphi')^{-1}\varphi''' + (\varphi''/\varphi')^2) .$$

Using Eqs.A.19-A.21 we see that

$$R/6 = (2\kappa_1 t + \kappa_2 + 1)(6\kappa_1 t + 3\kappa_2 + 2)\xi^2 + 2\kappa_1\xi . \qquad \text{Eq.A.31}$$

For the case where $\delta = 0$ we know that $\xi$ blows up as we approach $t = -\kappa_2/(2\kappa_1)$, which is not actually in the domain of the solution for $\varphi$ in this case. Nevertheless, Eq.A.31 tells us that R increases without bound as we approach that moment.

For the case where $\delta < 0$ we have the three solutions $\varphi_L$, $\varphi_M$ and $\varphi_R$, where $\xi$ is unbounded on the domain of each of these functions. Hence R is also unbounded on the domain of each of these solutions to $H = $ constant.

Lastly, for the case where $\delta > 0$, this solution is defined for all time with $\xi$ well-behaved. Hence due to Eq.A.31, R is well-defined and bounded. This case gives rise to a fairly tame universe in which $\varphi' = 0$ when $t = -\infty$, and then $\varphi'$ gradually increases to a maximum before returning to 0 at $t = \infty$.

In concluding this appendix I would like to say that in view of how difficult it is to solve non-linear third-order ordinary differential equations in terms of elementary functions, it is amazing that we were able to do just that for the equations $H_{2,4} = 0$, and $H$ = constant.  Recall that in the case of scalar-tensor field theories, the Lagrangian $L_{2,4}$ that we have been using, gives rise to field equations in which the speed of light is equal to the speed of gravitational waves, while that is not the case for $L_{2,5}$, and I have no idea of the relationship between these two quantities for the Lagrangian $L$ that generates $H$. That adds even more interest to our solutions to $H_{2,4} = 0$.

## Appendix B: The Multifurcation of Time and Multiverses

Back in the early 1970's I encountered a very interesting non-Hausdorff, 1-dimensional manifold, which I thought might prove useful to me some day. This manifold occurred as an exercise in Brickell & Clark [20] (*see,* problem 3.2.1 on page 40 of [20]). I shall presently describe a slight modification of this manifold, and then show how this manifold can be generalized to build a multiverse consisting of all the universes generated as solutions to $H_{2,4}=0$, of the form given in Eq.4.2.

Consider the subset $T$ of $\mathbb{R}^2$ defined by $T := \underset{\zeta \in \mathbb{R}}{\cup} U_\zeta$, where

$U_\zeta := \{ (t,0) \mid t<0 \} \cup \{ (t,\zeta) \mid t \geq 0 \}$ .

$\forall \; \zeta \in \mathbb{R}$ we define a 1-dimensional chart $t_\zeta \colon T \to \mathbb{R}$ with domain $U_\zeta$ by: $t_\zeta((t,0)) := t$, and $t_\zeta((t,\zeta)) := t$.  If $\zeta, \eta \in \mathbb{R}$, then $t_\zeta$ and $t_\eta$ are clearly $C^\infty$ related. The collection of charts $C(T)$ $:= \{t_\zeta\}_{\zeta \in \mathbb{R}}$ defines a $C^\infty$ atlas of $\boldsymbol{T}$ into $\mathbb{R}$, and hence $\boldsymbol{T}$ is a differentiable manifold of dimension 1 with its $C^\infty$ structure determined by $C(\boldsymbol{T})$. ( I am using Brickell & Clark's [20] definition of a differentiable manifold here, which does not require $\boldsymbol{T}$ to be endowed with a topological structure before the charts are defined.) The domains of the charts of the complete atlas of $\boldsymbol{T}$ determined by $C(\boldsymbol{T})$ determines a basis for a topology on $\boldsymbol{T}$ which we shall take as the manifold topology. In terms of this topology $\boldsymbol{T}$ is connected and satisfies the $T_1$ separation axiom. However, if $\zeta \neq \eta$ then it is impossible to separate the points $(0,\zeta)$ and $(0,\eta)$ by disjoint open sets. Thus the manifold topology of $\boldsymbol{T}$ is non-Hausdorff. Since $\mathbb{R}$ is uncountable, the atlas $C(\boldsymbol{T})$ is also uncountable. Due to Proposition 3.3.3 in [20], this implies that $\boldsymbol{T}$ does not admit a countable basis for its topology.

I define a function t: $\boldsymbol{T} \to \mathbb{R}$ by stipulating that t:= $t_\zeta$ on the domain of each chart $U_\zeta$. t is evidently well-defined and differentiable. We can think of t as the global time function on $\boldsymbol{T}$, which multifurcates at time t = 0, since $t^{-1}(0)$ is $\{ (0,\zeta) \mid \zeta \in \mathbb{R}\}$, while if $\gamma < 0$, then $t^{-1}(\gamma) = (\gamma,0)$. Let the symbol $\partial/\partial t$ denote the vector field on $\boldsymbol{T}$ defined by

$$\frac{\partial}{\partial t} := \frac{\partial}{\partial t_\zeta}$$

on the domain $U_\zeta$ of each chart $t_\zeta$. Evidently $\partial/\partial t$ is a well defined smooth vector field, even though t is not a global coordinate for $\boldsymbol{T}$. Due to the non-Hausdorffness of $\boldsymbol{T}$ the maximal integral curves of $\partial/\partial t$ starting at points of the form $(\gamma,0)$, when $\gamma < 0$, are not unique, and split as they pass through time t=0.

I shall now describe a generalization of the 1-dimensional manifold $T$ that we shall require for our multiverse. This manifold will also be 1-dimensional, and will embody multifurcating time.

Let $P_n$ be a subset of $\mathbb{R}^n$. In our applications $P_n$ will comprise the parameters that characterize different vacuum solutions. We let $T(P_n)$ be the subset of $\mathbb{R}^{n+1}$ defined by $T(P_n) := \underset{p \in P_n}{\cup} U_p$, where if $p = (p_1, \ldots, p_n) \in P_n$ then

$$U_p := \{ (t,0,\ldots,0) \in \mathbb{R}^{n+1} | \, t<0 \} \cup \{ (t, p_1, \ldots, p_n) \, | \, t \geq 0 \} \, .$$

$\forall \, p \in P_n$, I define a 1-dimensional chart $t_p$ with domain $U_p$ by

$$t_p((t,0,\ldots,0)) := t \text{ and } t_p((t, p_1, \ldots, p_n)) := t \, .$$

Due to my previous remarks it should be evident that the collection of charts $C(T(P_n))$ := $\{ t_p \, | \, p \in P_n \}$ determines a $C^\infty$ structure of dimension 1 on $T(P_n)$. With this structure $T(P_n)$ is a differentiable manifold of dimension 1. In terms of the manifold topology $T(P_n)$ satisfies the $T_1$ separation axiom, but is non-Hausdorff. This manifold is connected and admits a countable basis for its topology if and only if $P_n$ is countable. I define the global time function t and vector field $\partial/\partial t$ on $T(P_n)$ in a manner analogous to how they were defined on $T$. Once again we see that time essentially fractures as we pass through t=0, but even so the manifold remains connected.

The underlying manifold for our multiverse will be obtained by choosing $P_n$

suitably, and then crossing $\boldsymbol{T}(P_n)$ with $\mathbb{R}^3$. To that end let $\alpha$, $\beta$, and $\gamma$ be positive real numbers chosen so that $(\alpha, \beta, \gamma) \in \boldsymbol{Q'}(^3/_2)$. Consequently

$$\varphi_1(t) := \gamma(t)^{3/2}, \quad 0 \leq t < t_1$$

$$\varphi := \quad \varphi_2(t) := \alpha e^{\beta t}, \qquad t_1 \leq t \leq t_2 \qquad\qquad\qquad \text{Eq.B.1}$$

$$\varphi_3(t) := -\gamma(2t_2 - t)^{3/2}, \, t_2 < t \leq 2t_2 \, ,$$

is a class $C^{\infty,1}$ solution to $H_{2,4} = 0$, with $t_1$ and $t_2$ characterized by $t_1 < t_2$, $\varphi_1'(t_1^-) = \varphi_2'(t_1)$, and $\varphi_2'(t_2) = \varphi_3'(t_2^+)$. (Note Eq.B.1 is Eq.4.2, and that $\varphi_2'(t_2) = \varphi_1'(t_2)$, if we extend the domain of $\varphi_1$ to $t \geq 0$.) We let $\boldsymbol{T}(\boldsymbol{Q'}(^3/_2))$ denote the 1-dimensional manifold generated by the parameter space $\boldsymbol{Q'}(^3/_2)$ in the manner described above. The underlying 4-dimensional manifold of our multiverse based upon the vacuum solutions of $H_{2,4} = 0$, given in Eq.B.1, is $MV(\boldsymbol{Q'}(^3/_2)) := \boldsymbol{T}(\boldsymbol{Q'}(^3/_2)) \times \mathbb{R}^3$. At this juncture the bijective relationship between the quotient set $FMV(\boldsymbol{Q'}(^3/_2))/\sim$ defined in Section 4, and $MV(\boldsymbol{Q'}(^3/_2))$ should become somewhat apparent. It is this connection which enables us to define a manifold structure on $FMV(\boldsymbol{Q'}(^3/_2))/\sim$ that turns it into a quotient manifold diffeomorphic to $MV(\boldsymbol{Q'}(^3/_2))$.

$MV(\boldsymbol{Q'}(^3/_2))$ can be endowed with geometry in the following manner. If $(\alpha, \beta, \gamma) \in \boldsymbol{Q'}(^3/_2)$, then the line element on the 4-dimensional slice of $MV(\boldsymbol{Q'}(^3/_2))$ which consists of $U_{(\alpha, \beta, \gamma)} \times \mathbb{R}^3$ is defined by

$$ds^2 = -dt^2 + 0(du^2 + dv^2 + dw^2) \, , \text{ for } t < 0, \text{ and}$$

$$ds^2 = -dt^2 + (\varphi')^2(du^2 + dv^2 + dw^2) \text{ , for } t{\geq}0 \text{ ,}$$

where $\varphi$ is given by Eq.B.1.  Due to our previous work we know that the metric tensor, g, on each slice of MV($\boldsymbol{Q'}(^3/_2)$), is $C^\infty$ except when $t = 0$, $t_1$ and $t_2$, where it is continuous but not differentiable.  The curvature invariant $R^{abcd}R_{abcd}$ for the $\varphi_1$ part of $\varphi$ is given by $^3/_2\, t^{-4}$, and hence blows up as $t{\rightarrow}0^+$.  This is consistent with each vacuum universe beginning explosively with a singularity at t=0.  This invariant also blows up for $\varphi_3$ as $t{\rightarrow}2t_2^-$, and so this universe also ends with a singularity.

It should be evident how one could use the parameter space of other families of solutions of $H_{2,4} = 0$, to construct multiverses.  The issue now would be are any of these multiverses more plausible than any other?  And, moreover, given any one multiverse, are there particular slices that are more probable to arise than other slices?

The multiverse idea that I have just discussed might give us some insight into why our universe is dominated by matter, and does not contain equal amounts of matter and antimatter.  Now, admittedly, the multiverses I have described are built from source-free solutions of the field equations.  But let us suppose that when matter is present we can still construct a multiverse of the form MV($P_n$) for some parameter space $P_n$ associated with the solutions.  When t<0 all of the matter in the multiverse will be in a single $\mathbb{R}^3$.  And let us suppose that this infinite sea of matter is comprised of "equal" amounts of matter and antimatter.  Then when t=0, and time multifurcates, each quantum of matter is forced to go into some slice of the multiverse. We can only speculate on how particles make the decision where to go. But in any case, it would seem highly unlikely that after t=0 all of the particles in the various slices of $MV(P_n)$ are equally balanced between matter and antimatter. So during the period of time near $t=0^+$ one would expect that there would be a lot of annihilation between matter and antimatter, leaving a preponderance of one or the other as time evolves. One problem with this scenario (and I am sure that there are probably many more), is that when t<0, the original sea of equal amounts of matter and antimatter was an electrically neutral mixture of all different types of quarks and leptons. Then when these particles start spilling into the slices of $MV(P_n)$ the resulting mix would have a net electrical charge after all the matter and antimatter that can annihilate near $t=0^+$ does so. Our universe does not seem to possess a net charge (although that net charge could be small). One way to circumvent this "charge problem" is to assume that when t<0, the universe is only filled with equal numbers of hydrogen and anti-hydrogen atoms essentially doing nothing (don't ask me why they are not busy annihilating each other–perhaps there are no forces when t<0, but then what is holding the hydrogen and anti-hydrogen atoms together). Then when time multifurcates at t=0, these electrically neutral atoms start filling up the slices of the multiverse, leading to electrically neutral individual universes in which either matter or antimatter dominates.

In concluding this appendix I would like to mention that the multiverses constructed here are very different from the "traditional" multiverses, as described in Guth [19], which originate from false vacuum states of the inflaton field.

## Bibliography

[1]    P. Finsler, "über Kurven und Flächen in Allgemeinen Räumen," Ph.D. thesis, University of Göttingen, Göttingen, Germany, 1918.

[2]    H. Rund, "The Differential Geometry of Finsler Spaces," Springer-Verlag, 1959.

[3]    S.-S. Chern, D. Bao & Z. Shen, "An Introduction to Riemannian-Finsler Geometry," Springer-Verlag, 2000.

[4]    A. Bejancu & H. R. Farran, "Geometry of Pseudo-Finsler Submanifolds," Springer Science & Business Media B.V., 2000.

[5]    G. W. Horndeski, "Invariant Variational Principles and Field Theories," Ph.D. thesis, University of Waterloo, Waterloo, Ontario, 1973.

[6]    R. Adler, M. Bazin & M. Schiffer, "Introduction to General Relativity," McGraw-Hill Book Company, 1965.

[7]    G. W. Horndeski, "Second-Order Scalar-Tensor Field Theories in a Four-Dimensional Space," Inter. J. of Theo. Phys. **10** (1974), 363-384.

[8]    C. Deffayet, X. Gao, D. A. Steer & G. Zahariade, "From k-essence to generalized

Galileons," Phys. Rev. D **84**, 064039 (2011); arXiv.org/abs/1103.3260, March, 2011.

[9]   T. Kobayashi, M. Yamaguchi & J. Yokoyama, "Generalized G-inflation: Inflation with the most general second-order field equations," Prog. Theor. Phys. **126**, (2011) 511-526; arXiv.org/abs/1105.5723, September, 2011.

[10]  M. Ostrogradsky, "Memories sur les equations differentielles relatives au problems des isoperimetrics," Mem. Ac. St. Petersburg **4**, (1850), 385.

[11]  R. P. Woodard, "The Theorem of Ostrogradsky," arXiv.org/abs/1506.02210, August, 2015.

[12]  T. Baker, E. Bellini, P. G. Ferreira, M. Lagos, J. Noller & I. Sawicki, "Strong constraints on cosmological gravity from GW170817 and GRB170817A," Phys. Rev. Lett. **119**, 251301 (2017); arXiv.org/abs/1710.06394, October, 2017.

[13]  P. Creminelli & F. Vernizzi, "Dark Energy after GW170817 and GRB170817A," Phys. Rev. Lett. **119**, 251302 (2017); arXiv.org/abs/1710.05877, December, 2017.

[14]  J. Sakstein & B. Jain, "Implications of the Neutron Star Merger GW170817 for Cosmological Scalar-Tensor Theories," Phys. Rev. Lett. **119**, 251303 (2017); arXiv.org/abs/1710.05893, December, 2017.

[15]  J. M. Ezquiaga & M. Zumalácarregui, "Dark Energy after GW170817: dead ends

and the road ahead," Phys. Rev. Lett **119**, 251304 (2017); arXiv.org/abs/1710.05901, November, 2017.

[16]  G. W. Horndeski, "Conformally Invariant Scalar-Tensor Field Theories in a 4-Dimensional Space," arXiv.org/abs/1706.04827, June, 2017.

[17]  G. W. Horndeski, "Conformally Invariant Scalar-Vector-Tensor Field Theories Consistent with Conservation of Charge in a Four-Dimensional Space," Fund. J. of Modern Phys. **11**, 93-133 (2018); arXiv.org/abs/1801.02490, January, 2018.

[18]  G. W. Horndeski, "Conservation of Charge and the Einstein-Maxwell Field Equations," J. Math. Phys. **17**, 1980 (1976).

[19]  A. H. Guth, "The Inflationary Universe: The Quest for a New Theory of Cosmic Origins," Basic Books, 1998.

[20]  F. Brickell. & R.S. Clark, "Differentiable Manifolds: An Introduction," Van Nostrand Reinhold Company, 1970.

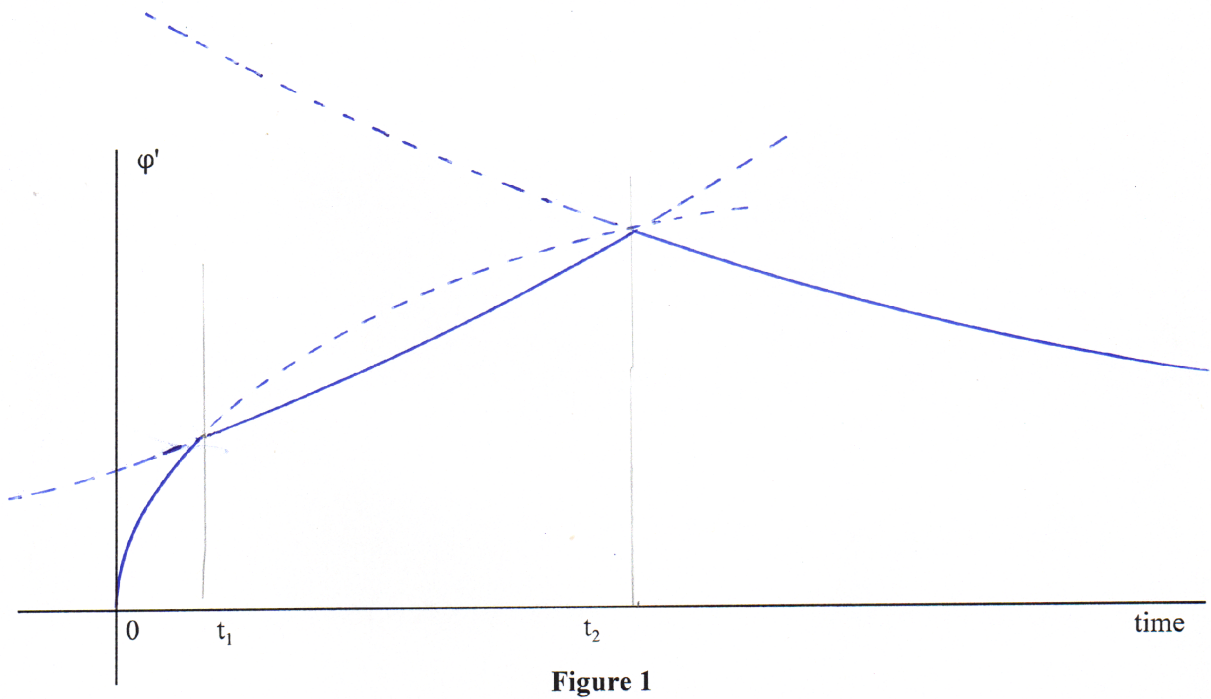

**Figure 1**

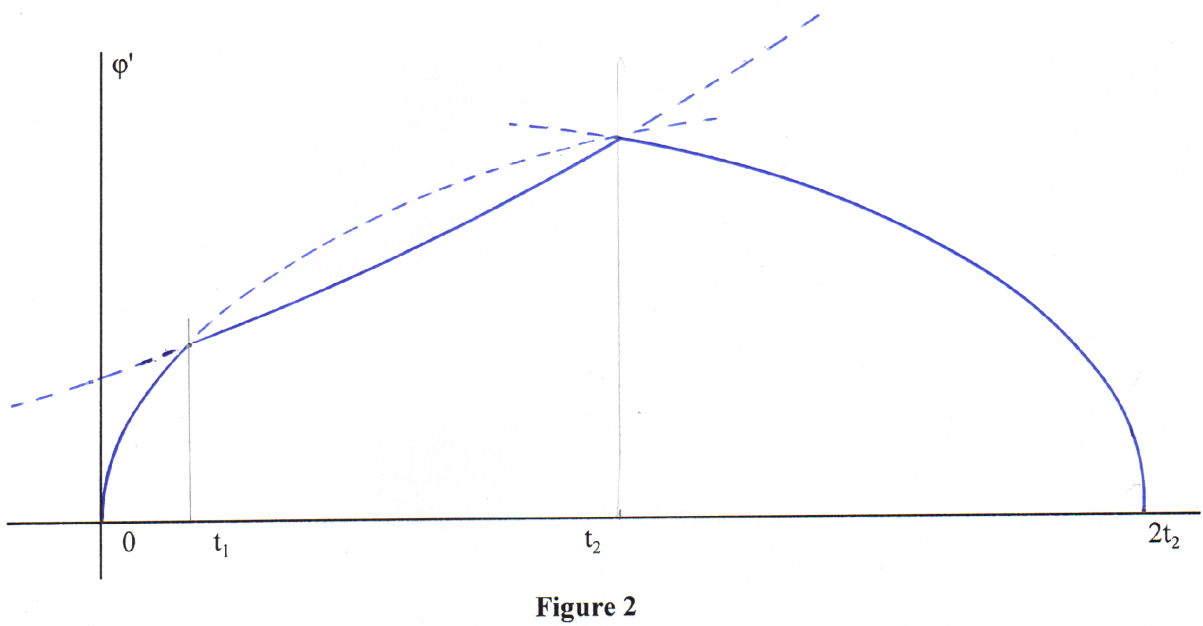

**Figure 2**

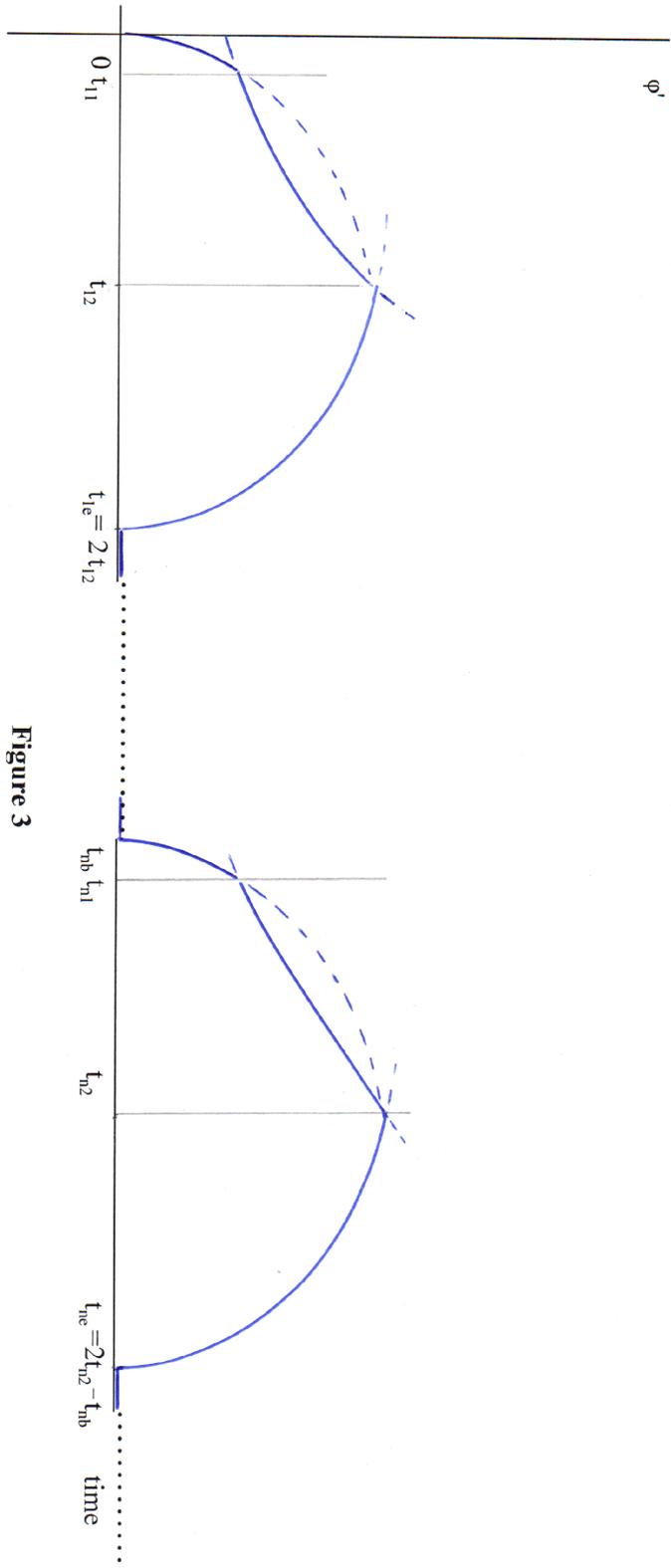

**Figure 3**

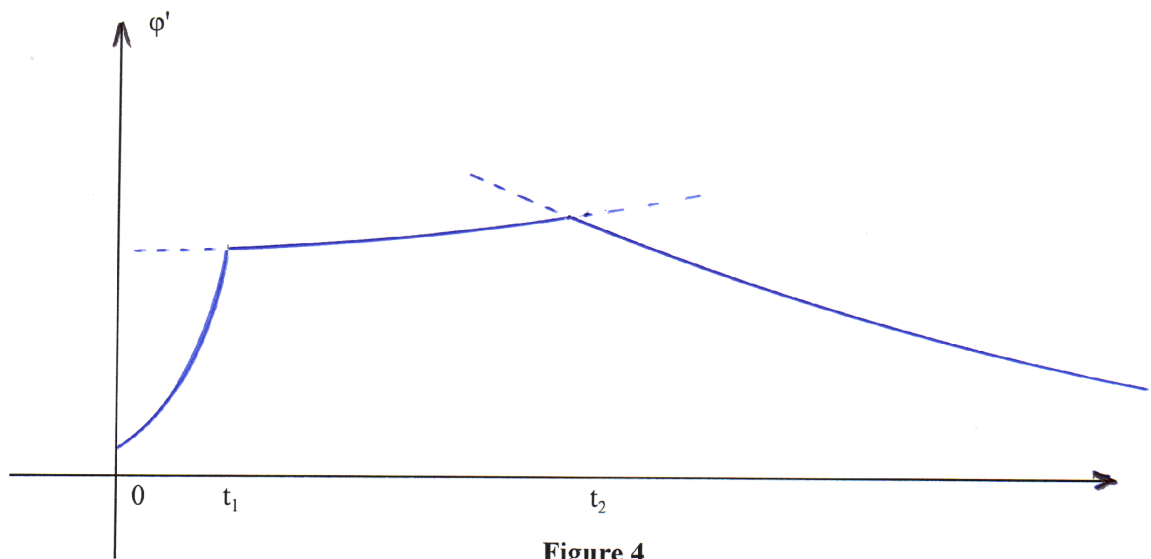

**Figure 4**